\documentclass{aa}
\usepackage{graphicx}
\usepackage{natbib}
\usepackage{txfonts}
\bibpunct{(}{)}{;}{a}{}{,}
\newcommand{\kms}{\,km\,s$^{-1}$}
\newcommand{\cmd}{\,cm$^{-2}$}
\newcommand{\cmt}{\,cm$^{-3}$}
\newcommand{\eps}{$\rm\,erg\,cm^{3}\,s^{-1}$}
\newcommand{\myr}{\,$M_{\sun}\,{\rm yr}^{-1}$}
\newcommand{\ecs}{$\rm\,erg\,cm^{-2}\,s^{-1}$}
\newcommand{\ecsa}{$\rm\,erg\,cm^{-2}\,s^{-1}\,\AA^{-1}$}
\newcommand{\ha}{H$\alpha$~}
\newcommand{\hb}{H$\beta$~}
\def\V{\rm {\scriptsize V}}
\def\I{\rm {\scriptsize I}}
\begin{document}
\title{Structure of the hot object in the symbiotic prototype Z\,And \\
       during its 2000-03 active phase}

\author{A.~Skopal\thanks{Visiting Astronomer: Capodimonte Astrophysical 
                         Observatory} %, Naples, Italy}
        \inst{1}
        \and
         A. A.~Vittone
        \inst{2}
        \and
         L.~Errico
        \inst{2}
        \and
         M.~Otsuka
        \inst{3}
        \and
         S.~Tamura
        \inst{4}
        \and
         M.~Wolf
        \inst{5}
        \and
         V. G.~Elkin
        \inst{6}
}
\institute{Astronomical Institute, Slovak Academy of Sciences,
           059\,60 Tatransk\'{a} Lomnica, Slovakia 
       \and
           INAF Osservatorio Astronomico di Capodimonte, 
           via Moiariello 16, I-80\,131 Napoli, Italy
       \and 
           Okayama Astrophysical Observatory, NAOJ, Kamogata, 
           Okayama 719-0232, Japan
       \and
           Astronomical Institute, Tohoku University, 
           Sendai 980-8578, Japan 
       \and
           Astronomical Institute, Charles University Prague, 
           CZ-18000 Praha 8, V Hole\v sovi\v ck\'ach 2, 
           Czech Republic
       \and
           Centre for Astrophysics, University of Central Lancashire 
           Preston PR1~2HE, United Kingdom}

\date{Received; accepted}

\abstract
  {}
%Aims
  {To investigate structure of the hot object in the symbiotic 
   prototype Z\,And during its major 2000-03 active phase.}
%Methods
  {Analysis of the far ultraviolet, optical low- and high-resolution 
   spectroscopy and $UBVR$ photometry. Reconstruction of the 
   spectral energy distribution (SED) during the outburst. 
   The Raman scattering process.}
%Results
  {At the initial stages of the outburst the hot object was 
   characterized by the two-temperature spectrum (a warm stellar 
   radiation and a strong nebular emission) with signatures of 
   a mass-outflow at moderate ($\sim$\,100$\div$200\kms) and 
   very high ($\approx$\,1000$\div$2000\kms) velocities. 
   The corresponding structure of the hot object consists of 
   an optically thick, slowly-expanding disk-like material 
   encompassing the accretor at the orbital plane and a fast 
   optically thin wind over the remainder of the star. 
   The disk-like shell persisted around the central star 
   until 2002 August as was indicated by the eclipse effect. 
   Then a significant dilution of the optically thick 
   material and evolution of a fast wind from the hot star, 
   concentrated more at the orbital plane, were detected. 
   A striking similarity of [\ion{Fe}{vii}]\,$\lambda$6087 and 
   Raman $\lambda$6825 profiles at/after the dilution of the 
   disk suggested their origin within the interaction zone 
   where the winds from the binary components collide.}
%  {A multiple mass ejection from the active object was indicated 
%   on the rise to the optical maximum. 
  {}
\maketitle

\section{Introduction}
%^^^^^^^^^^^^^^^^^^^^^

Z\,And is considered as a prototype of the class of symbiotic
stars. The binary composes of a late-type, M4.5\,III, giant
and a white dwarf accreting from the giant's wind on the 758-day
orbit \citep[e.g.][]{nv89,mk96}. More than 100 years of 
monitoring Z\,And (first records from 1887) demonstrated 
an eruptive character of its light curve (LC). It displays 
several active phases, during which fluctuations ranges in 
amplitude from 
a few tenths of a magnitude to about 3 magnitudes. During 
the strongest outbursts (1915, 1940, 1960, 1985, 2000) the 
star's brightness reached values around $m_{\rm pg} \sim$\,9 
and $m_{\rm vis} \sim$\,8.5\,mag with the recurrence time of 
15 to 25 years (see Fig.~1). 
Major outbursts are characterized by a decrease in the 
effective temperature of the hot object and the fluxes 
of some high ionization emission lines. 
Analyzing previous large eruptions, \cite{mk96} found that 
an A--F absorption spectrum characterized the 1940 optical 
maximum and during the 1960 outburst the system resembled 
an F0--2 supergiant. They estimated a roughly constant 
luminosity and a factor of 10 lower temperature, $T_{\rm h}$, 
for the hot object during eruptions with respect to quantities 
from quiescence. As concerns to the 1985 eruption, 
\cite{mk96} supported the decrease in $T_{\rm h}$ also by 
a decline of Raman scattered \ion{O}{vi} lines at 
$\lambda$6825 and $\lambda$7088. 
This outburst was studied in detail by \cite{fc+95}. 
By means of the ultraviolet spectroscopy and radio 
observations, they indicated a considerably cooler UV 
continuum with an additional attenuation below 1300\,\AA\ 
accompanied by a decline in the radio flux 
during the 1985 optical maximum. In the line spectrum they 
measured a broadening of all emission lines with signatures 
of P-Cygni type of profiles for strong resonance lines. 
\cite{fc+95} interpreted these features as a result of the 
ejection of an optically thick shell at moderately high 
velocity of 200--300\kms. 
Radio observations also provided evidences of a mass-outflow 
from the active star. At the initial stage of both 1985 and 2000 
outbursts the radio flux dropped below values from quiescence, 
then recovered along the optical decline and reached close to 
the quiescent level or above it \citep{fc+95,b+04}. For 
the 1984-85 activity, \cite{fc+95} explained the anticorrelation 
between the visual and far-UV LCs by the extinction of UV radiation 
from the central source by a false photosphere 
produced by the ejection of a shell of dense material. They 
compared the subsequent rise of the radio emission with a model, 
in which the ionizing source is turned on instantaneously 
at the center and an ionization front propagates out through 
the wind from the hot star. 
For the 2000-03 active phase, \cite{b+04} revealed a jet-like 
extension at 5\,GHz on the 2001 September MERLIN image. They 
associated this transient feature to a blob of matter ejected 
at the beginning of the optical outburst, emitting most 
probably through thermal bremsstrahlung of ionized hydrogen gas. 

On 2000 September 1st Z\,And entered the recent major outburst 
\citep{sk+00}. It reached a maximum brightness around the mid 
of 2000 December ($U\,\sim$\,8.4, $B\,\sim$\,9.3 and 
$V\,\sim$\,8.8\,mag) following a gradual decrease to quiescent 
values at/after the mid of 2003 \citep{sk+04}. 
In the summer of 2002, \cite{sk03a} observed a narrow minimum 
in the LC at the position of the inferior conjunction of the 
giant. He associated this effect to the eclipse of the active 
object by the giant and determined the orbital inclination of 
the system to $\ga\,76^{\circ}$. 
\cite{sok+05} observed the 2000-03 outburst in the X-rays, 
far ultraviolet, optical and radio. They focused on the nature 
of the outburst, and described observational evidence for 
a new kind of the outburst that involves both an accretion-disk 
instability and an thermonuclear shell burning on the surface 
of an accreting white dwarf. 
%
%summarized first observational results from the 
%initial stage of the outburst. Particularly, their $FUSE$ 
%({\em Far Ultraviolet Spectroscopic Explorer}) spectra 
%indicated a large amount of expanding cool gas, which only 
%partially covers the source of FUV emission and their $X$-ray 
%observations interpreted as a result of collisions between 
%the outburst ejecta and the red giant wind. 

In this paper we present and analyze our original spectroscopic 
and photometric observations complemented by the archival spectra 
obtained with the {\em Far Ultraviolet Spectroscopic Explorer} 
($FUSE$) carried out during the 2000-03 major outburst of Z\,And. 
%
% - reconstruction of the hot object structure 
%
In Sect.~3 we analyze the data. We model the SED (Sect.~3.3) 
and examine the Raman scattering process (Sect.~3.4). In 
Sect.~4 we summarize main results to reconstruct a basic 
structure of the hot object and the surrounding environment 
during the outburst. 
%
%==================================================|
%------------- Table 1: Used spectra --------------|
%==================================================|
%
\begin{table}
\caption[]{Log of spectroscopic observations}
% \begin{flushleft}
\begin{center}
\begin{tabular}{ccccc}
\hline
\hline
 Date & Julian date    & Phase$^{\star}$ & Region  & Observatory  \\
      & JD~2\,4...     &                 &  [nm]   &              \\
% \noalign{\smallskip}
\hline
% \noalign{\smallskip}
02/11/00 & 51851.27 &0.143 &   409.2 - 494.8 & SAO    \\
04/11/00 & 51853.34 &0.146 &   413.2 - 495.0 & SAO    \\
05/11/00 & 51854.37 &0.147 &   413.8 - 495.0 & SAO    \\
05/11/00 & 51854.39 &0.147 &   482.0 - 563.1 & SAO    \\
16/11/00 & 51865.00 &0.161 &    90.5 - 118.8 & FUSE   \\
27/11/00 & 51876.00 &0.176 &    90.5 - 118.8 & FUSE   \\
10/12/00 & 51889.41 &0.194 &   459.5 - 909.2 & Asiago \\
11/12/00 & 51890.35 &0.195 &   459.5 - 909.2 & Asiago \\
12/12/00 & 51891.30 &0.196 &   459.5 - 909.2 & Asiago \\
15/12/00 & 51894.00 &0.200 &    90.5 - 118.8 & FUSE   \\
05/01/01 & 51914.92 &0.227 &   580.0 - 700.0 & OAO    \\
08/01/01 & 51917.90 &0.231 &   450.0 - 570.0 & OAO    \\
27/05/01 & 52057.26 &0.415 &   600.0 - 700.0 & OAO    \\
22/07/01 & 52113.30 &0.489 &    90.5 - 118.8 & FUSE   \\
30/09/01 & 52183.30 &0.582 &    90.5 - 118.8 & FUSE   \\
10/02/02 & 52315.93 &0.757 &   580.0 - 700.0 & OAO    \\
05/07/02 & 52461.50 &0.949 &    90.5 - 118.8 & FUSE   \\
07/08/02 & 52494.25 &0.992 &   580.0 - 700.0 & OAO    \\
22/10/02 & 52570.50 &0.093 &    90.5 - 118.8 & FUSE   \\
02/02/03 & 52673.45 &0.229 &   580.0 - 700.0 & OAO    \\
31/07/03 & 52852.50 &0.465 &   580.0 - 700.0 & OAO    \\
04/08/03 & 52856.80 &0.471 &    90.5 - 118.8 & FUSE   \\
02/09/03 & 52885.50 &0.509 &   580.0 - 700.0 & OAO    \\
13/11/03 & 52957.43 &0.604 &   625.8 - 677.0 & Ond\v rejov\\
18/12/03 & 52992.28 &0.650 &   625.8 - 677.0 & Ond\v rejov\\
\hline
\end{tabular}
\end{center}
$^{\star}$~$JD_{\rm Min} = 2\,414\,625.2 + 757.5\times E$ 
           \citep{sk98}
\end{table}

\section{Observations and reductions}
%^^^^^^^^^^^^^^^^^^^^^^^^^^^^^^^^^^^^

\subsection{Optical spectroscopy}
%^^^^^^^^^^^^^^^^^^^^^^^^^^^^^^^^

Our spectroscopic observations were taken during the recent 
major 2000-03 active phase at different observatories. 
%
%and phases of evolution in the LC. 

% SAO RAN:
Prior to the maximum of the star's brightness a low-dispersion 
spectroscopy was secured at the Special Astrophysical Observatory 
of the Russian Academy of Sciences (SAO in Table~1) with the long 
slit {\small UAGS} spectrograph equipped with a fast Schmidt camera 
and CCD detector (530$\times$580\,pixels of 18$\times$24\,$\mu$m 
size) mounted at the Cassegrain focus of the Zeiss 1-m telescope. 
Linear dispersion was 1.5 and 3.1\,\AA\,pixel$^{-1}$ in the blue 
and red part of the spectrum, respectively. 
Basic reduction including extraction of 1D-spectra from CCD 
images and wavelength calibration were made by using the 
software-package of \cite{vlasyuk}. 

% Asiago:
At the maximum of the star's brightness a high-dispersion 
spectroscopy was secured at the Asiago Astrophysical 
Observatory (OAC) with the {\small REOSC} Echelle Spectrograph 
({\small RES}) equipped with a Thompson {\small THX31156} 
UV-coated CCD detector 
(1024$\times$1024 pixels of 19\,$\mu$m size) mounted at 
the Cassegrain focus of the 1.82-m telescope at Mt.~Ekar. 
Two very different exposures of 120 and 3\,600\,s were 
applied to obtain well defined H$\alpha$ profile and 
the underlying continuum. The {\small RES} orders were 
straightened through the software
developed at the Astronomical Observatory of Capodimonte in
Napoli. Thereafter, the spectroscopic data were processed by
the {\small MIDAS} software package.
Dispersion of the {\small RES} was 10\,\AA\,mm$^{-1}$ 
(0.19\,\AA\,pixel$^{-1}$) at H$\alpha$. 
The spectra obtained during each night cover the wavelength 
range of 459.5 -- 909.2\,nm.

% Okayama: 
After the maximum and the following transition to quiescence 
a high-dispersion spectroscopy was secured at the Okayama 
Astrophysical Observatory (OAO) with the HIgh Dispersion 
Echelle Spectrograph \citep[{\small HIDES},][]{i99} at the f/29 
coud\'{e} focus of the 1.88-m telescope. Dimension of 
the CCD ({\small EEV} 42-80) was 2048$\times$4096 pixels of 
13.5\,${\rm \mu m}$ size. The spectral resolving power was 
50\,000 or 68\,000. The dispersion was 
$\sim$0.02\,{\AA}\,pixel$^{-1}$ ($\sim$\,1.2\,km\,s$^{-1}$) 
and 
$\sim$0.03\,{\AA}\,pixel$^{-1}$ ($\sim$\,1.4\,km\,s$^{-1}$)
in the region of 450 -- 570 and 580 -- 700\,nm, respectively. 
Bias and instrumental flats were also measured and taken into 
account in the usual way. The reduction and analysis was 
performed with the {\small IRAF}-package software and our own 
code to reduce spectra obtained with {\small HIDES}. 
For example, the scattered light from the instrument was 
also removed. 
The Th-Ar comparison spectrum was fitted with a 3rd-order 
polynomial function, which reproduced the wavelength 
calibration to better than 10$^{-3}$\,\AA. 

% Ondrejov: 
At the post-outburst stage (the end of 2003) a high-resolution 
spectroscopy between 6258 and 6770\,\AA\ was taken at the 
Ond\v{r}ejov Observatory using the coude spectrograph of 2-m 
reflector and the {\small BROR} CCD camera with the SITe 
800$\times$2030\,pixels chip. The resolution power at 
the H$\alpha$ region was 13\,000. 
Basic treatment of the spectra was done by using 
the {\small IRAF}-package software. 

A correction for heliocentric velocity was applied to all 
spectra. The journal of spectroscopic observations is 
in Table~1. 

\subsubsection{Flux calibration of the spectra}
%^^^^^^^^^^^^^^^^^^^^^^^^^^^^^^^^^^^^^^^^^^^^^^

We converted arbitrary flux units of our spectra to fluxes 
in \ecsa\ with the aid of the simultaneous $UBVR$ photometry. 
Following points are relevant: 
  (i) We summarized photometric observations carried out 
(nearly-)simultaneously with the spectrum under consideration. 
  (ii) We calculated corrections $\Delta m$, which are due to 
emission lines in the spectrum, to get magnitudes corresponding 
to the continuum \citep[see][for details]{sk03b}. 
  (iii) Then we dereddened the continuum-magnitudes with 
$E_{\rm B-V}$ = 0.30 by using the extinction curve of 
\cite{c+89} and converted to fluxes according to the 
calibration of \cite{hk82}. 
   (iv) Finally, we determined the star's continuum by fitting 
the $UBVR$ fluxes with a 3-order polynomial function and scaled 
the measured spectrum to this fit. 

\subsection{Far ultraviolet spectroscopy}
%^^^^^^^^^^^^^^^^^^^^^^^^^^^^^^^^^^^^^^^^

The $FUSE$ instrument was described in detail by \cite{moos} 
and \cite{sahnow}. A brief useful description of the spectrograph 
is given in \cite{young}. 
%
%For the purpose of this paper we used 
%archival observations taken during the investigated 
%outburst of Z\,And (Table~1). 

The data were processed by the calibration pipeline version 
3.0.7, 3.0.8 and 2.4.1. We used the calibrated time-tag 
observations ({\small TTAG} photon collecting mode) taken 
through the large aperture (LWRS). 
Before adding the flux from all exposures we applied an 
appropriate wavelength shift relative to one so to get 
the best overlapping of the absorption features. Then we 
co-added spectra of individual exposures and weighted them 
according to their exposure times. Finally, we binned the 
resulting spectrum within 0.025\,\AA. 
In order to determine the wavelength scale of the spectra, 
first a correction for heliocentric velocity including 
that of the satellite was applied. Second we selected 
interstellar absorption lines for which wavelengths 
are known and fitted the corresponding observed features. 
For example, in the region of the LiF1A spectra we used 
\ion{N}{i}\,$\lambda$1003.372, 
\ion{Si}{ii}\,$\lambda$1020.699, 
\ion{A}{i}\,$\lambda$1048.218 
\citep[e.g.][Fig.~3 here]{re84}. 
An accuracy of such calibration is of $\pm$\,0.05\,\AA. 
The data retrieved from the $FUSE$ archive were processed 
mostly by own codes (AS). 
%
%- Reddening, extinction curve:

Dereddening the $FUSE$ spectra we found the extinction curve 
of \cite{c+89} to be valid only in the interval from the 
long-wavelength end of the spectra to about 1\,100\,\AA\ 
(open circles in Figs.~6 and 7; $E_{\rm B-V}$ = 0.30, 
$R_{\rm V}$ = 3.1). Beyond these wavelengths to 1\,000\,\AA\ 
dereddened fluxes enlarges by a factor of 1.5$\div$2 with respect 
to quantities required by our modeling the SED (crosses in 
Fig.~7). Note that the extinction curve in the 
far-ultraviolet (1\,250$\div$1\,000\,\AA) as suggested 
by Cardelli et al. (1986, Fig.~5 there) represents only an 
extrapolation and thus is more uncertain. As a result we used 
their mean extinction law only for $\lambda > 1\,100$\,\AA. 
The short-wavelength rest of the spectrum was rectified to 
our model SED (Fig.~6). 
%
%Finally, we note 
%Ly$\beta$ photons.
%The observed flux was in (major) part of a geocoronal nature,
%because of its large variability, up to 100\,\%, between
%measurements by LiF1A and SiC1A channels. Also the flux
%appearing in the adjacent MDRS aperture suggested it
%as a strong airglow emission. This concerns mainly to
%observations from the optical maximum.
%

\subsection{Broad-band photometry}
%^^^^^^^^^^^^^^^^^^^^^^^^^^^^^^^^^

Photometric observations were performed in the standard 
Johnson $UBV$ system with the $R_{\rm C}$ filter of the 
Cousins system. We used a single-channel photoelectric 
photometers mounted in the Cassegrain foci of 0.6-m 
reflectors at the Skalnat\'{e} Pleso and Star\'{a} 
Lesn\'{a} observatories. 
We converted our $R_{\rm C}$ magnitude to that of Johnson
system by using transformations of \cite{bessell}.
Z\,And was measured with respect to the comparison
SAO\,53150 (BD+47\,4192; $V$ = 8.99, $B - V$ = 0.41, 
$U - B$ = 0.14, $V - R$ = 0.16). 
Usually, 1-hour cycle contained about 10 to 20 individual 
differences between the target and the comparison.
This approach reduced the {\em internal} uncertainty of such 
the means to $\sim$0.01\,mag in the $V$, $B$ and $R$ bands, 
and up to 0.025\,mag in the U band. Individual 
measurements in a form of the table were already 
published by \cite{sk+04}. 

In this paper we dereddened our observations with 
$E_{\rm B-V}$ = 0.30 and scaled relevant parameters 
for a distance of 1.5\,kpc 
\citep[][and references therein]{sk05}. 
%
%==========================================|
%---------- Fig. 1: LCs of Z And  ---------|
%==========================================|
%
\begin{figure*}
\centering
\begin{center}
\resizebox{15cm}{!}{\includegraphics[angle=-90]{zhlc.epsi}}

\vspace*{3mm}
\resizebox{15cm}{!}{\includegraphics[angle=-90]{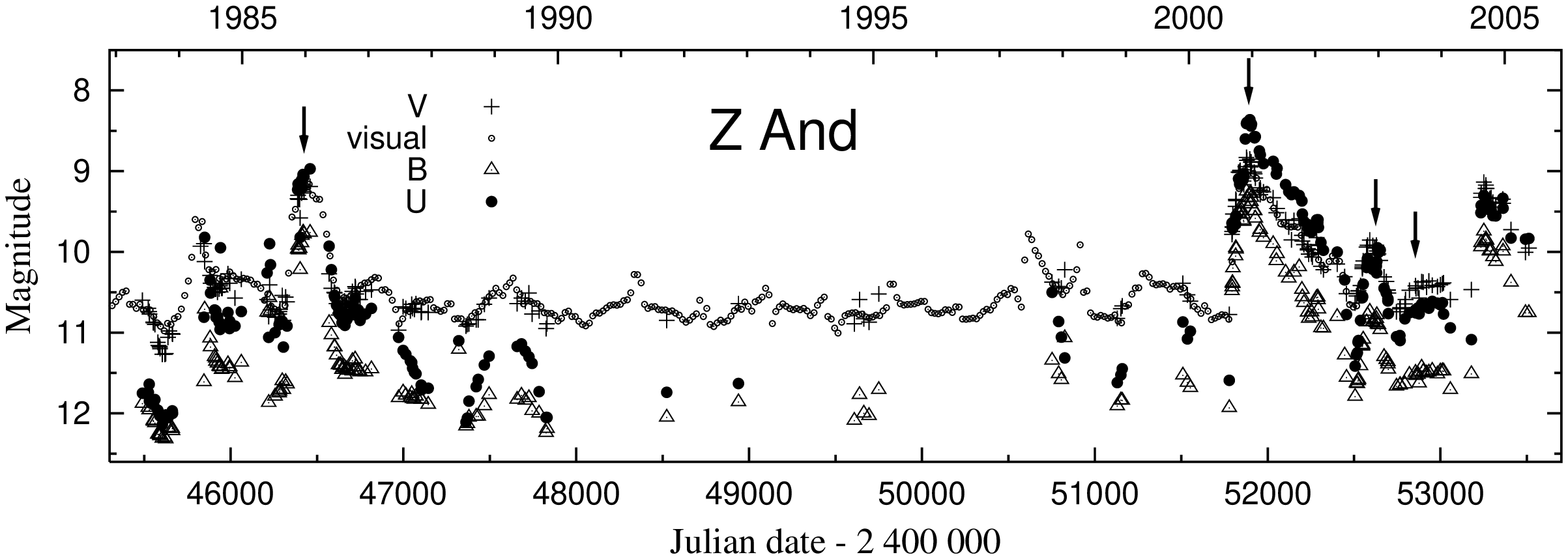}}

\vspace*{2mm}
\resizebox{15cm}{!}{\includegraphics[angle=-90]{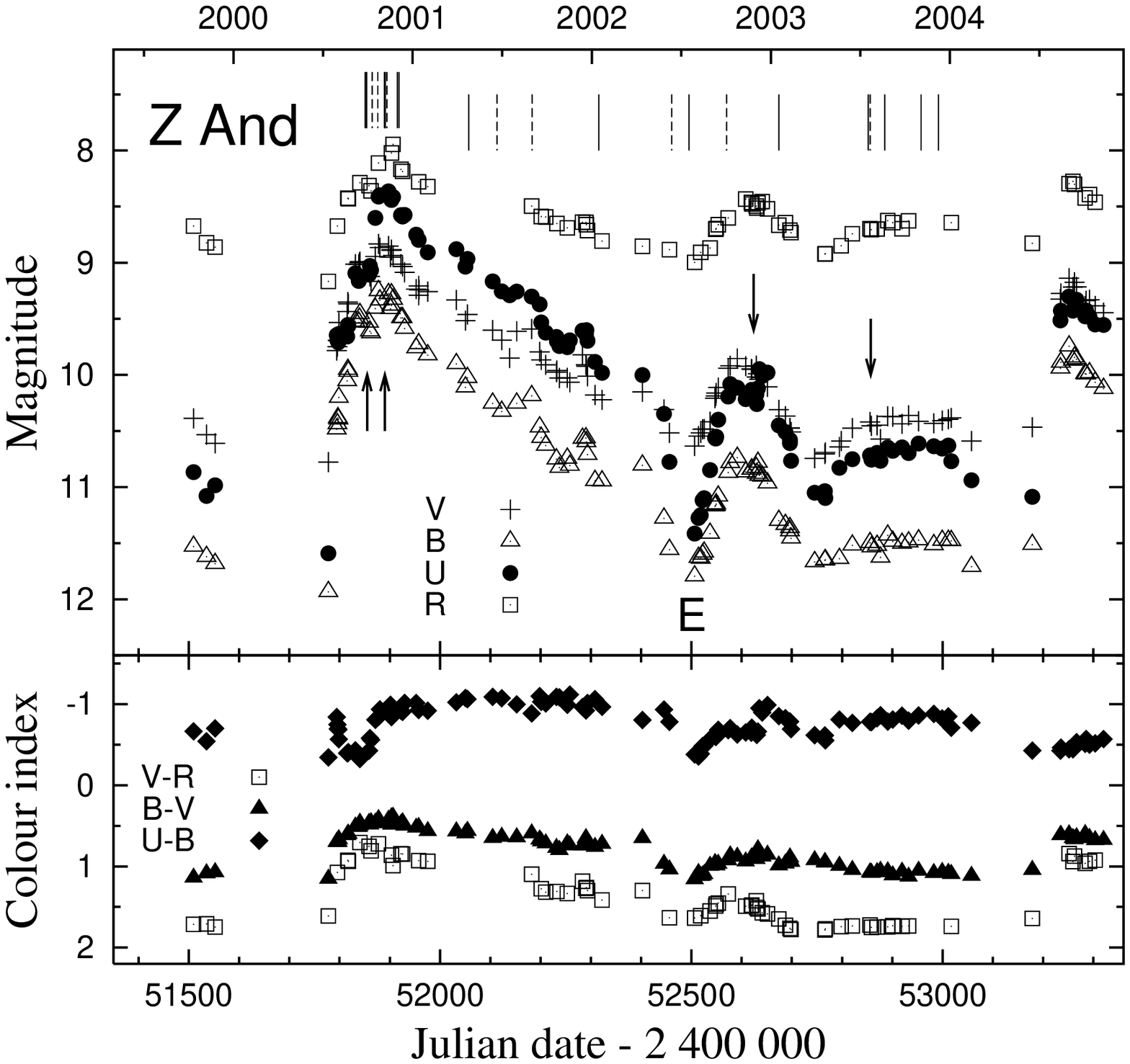}
                    \includegraphics[angle=-90]{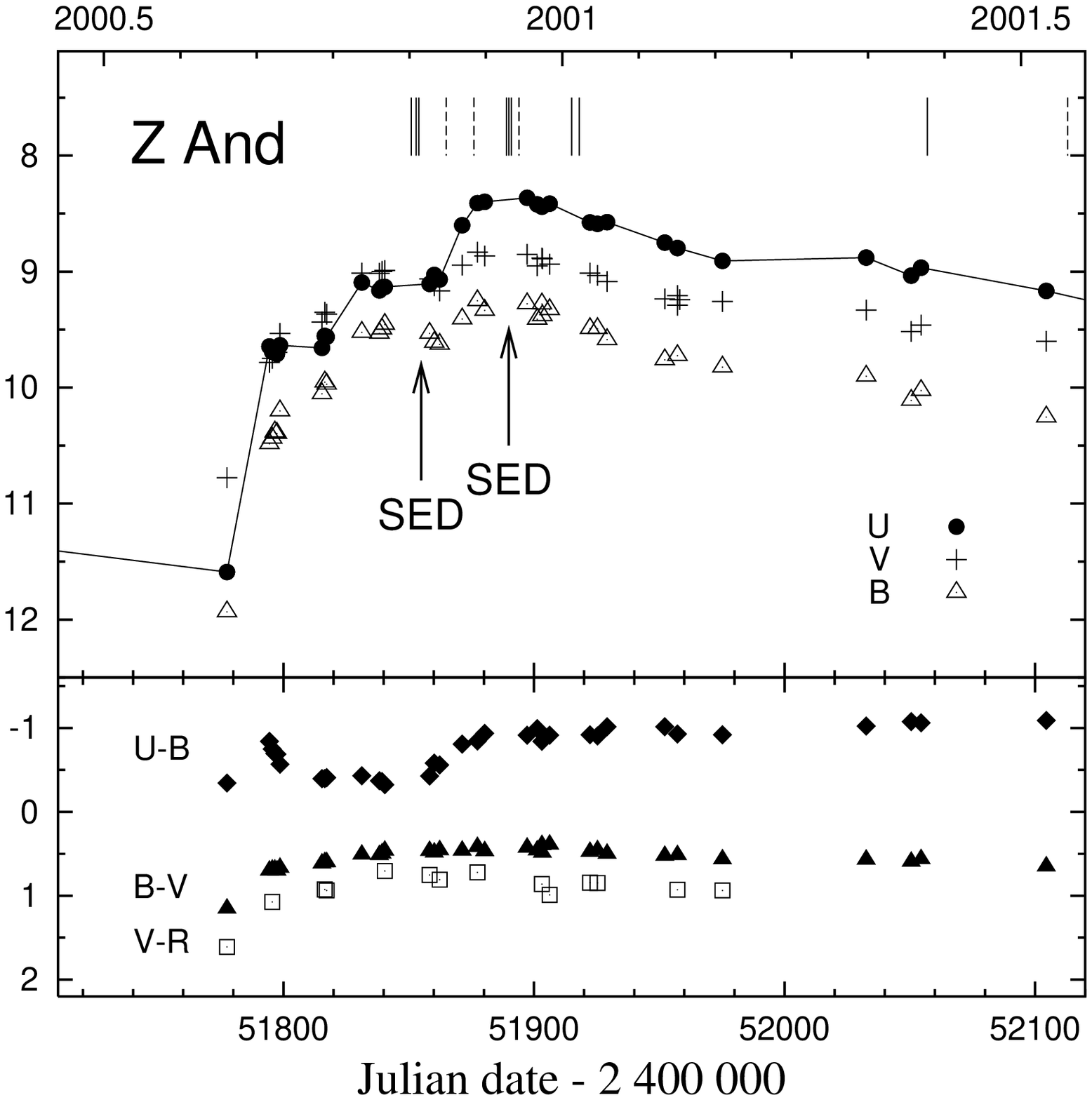}}
\caption[]{
Top: 
The historical 1895-2004 photographic/B-band/visual
LC of Z\,And. It is compiled from the data published by 
\cite{pg46}, \cite{r60}, \cite{mjal77}, \cite{bel85,bel92},
\cite{mk96}, \cite{sk+02b,sk+04} and smoothed visual AFOEV 
data. 
Middle:    
A part of the multicolour LC covering the recent two major 
outbursts (1984-86 and 2000-03).
Bottom: 
The $UBVR$ photometry from the major 2000-03 outburst. 
 Left panel shows the overall evolution in the LC. 
 The eclipse is denoted by {\sf E}.
The right panel shows the LC at the initial stage of the 
outburst -- note its stepped profile on the rise to the maximum. 
Optical/FUV spectroscopic observations are denoted by vertical 
full/dashed bars. The arrows mark positions at which we 
reconstructed the observed SED. 
          }
\end{center}
%\label{fig_1}
\end{figure*}
%
%
%========================================================|
%-- Fig. 2: Spectrum from the maximum  + HeI evolution --|
%========================================================|
%
\begin{figure*}
\centering
\begin{center}
\resizebox{\hsize}{!}{\includegraphics[angle=-90]{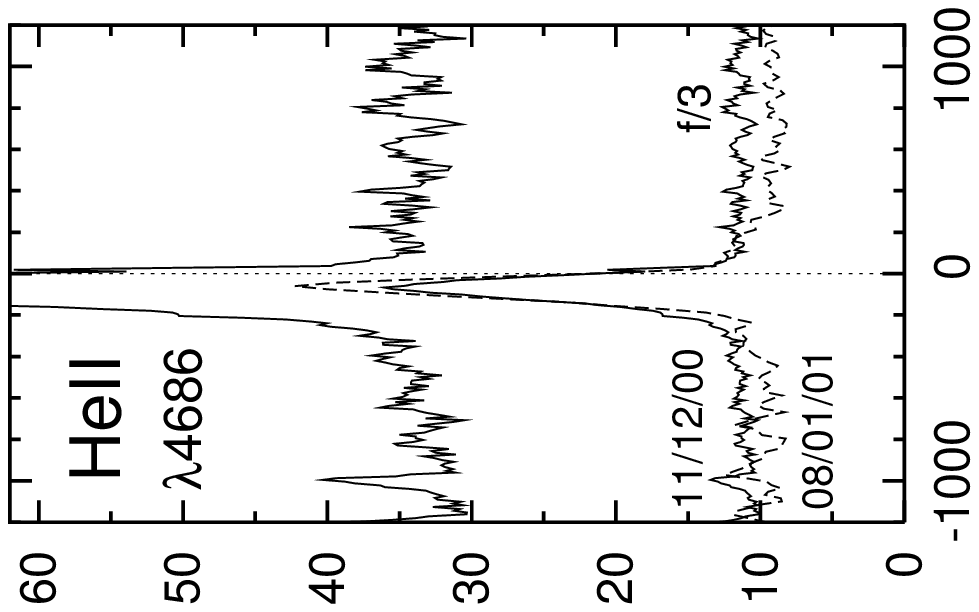}
                      \includegraphics[angle=-90]{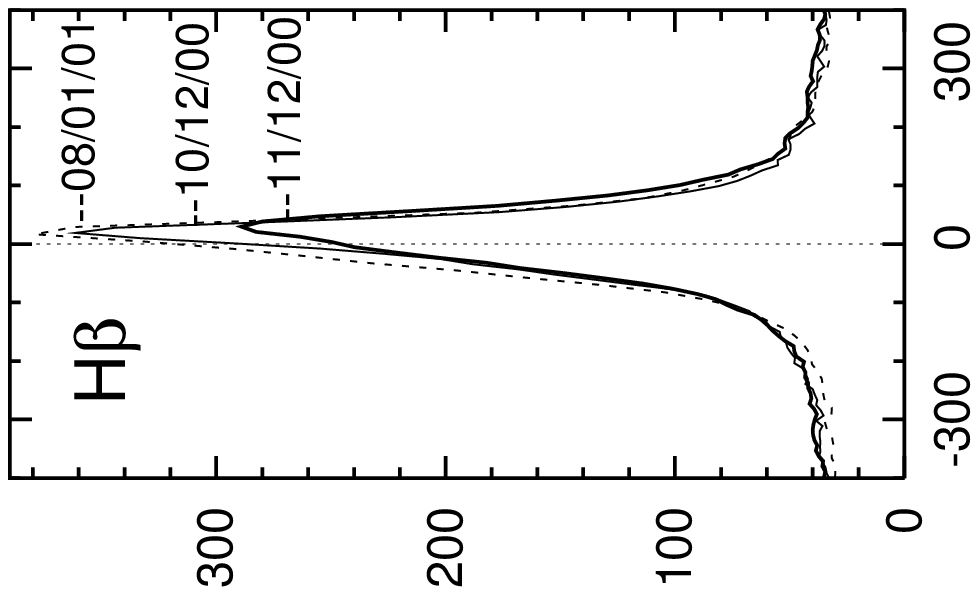}
                      \includegraphics[angle=-90]{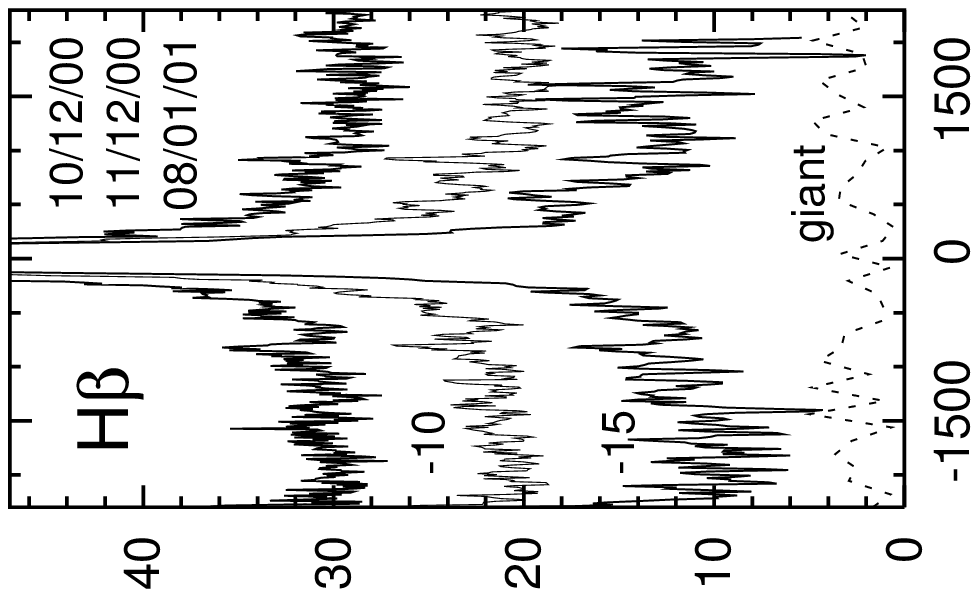}
                      \includegraphics[angle=-90]{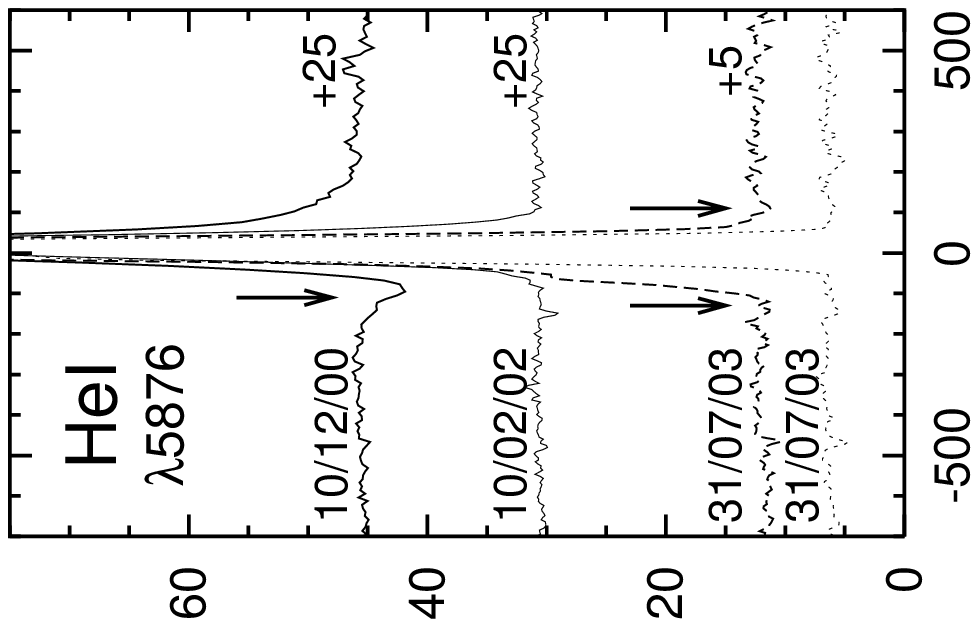}
                      \includegraphics[angle=-90]{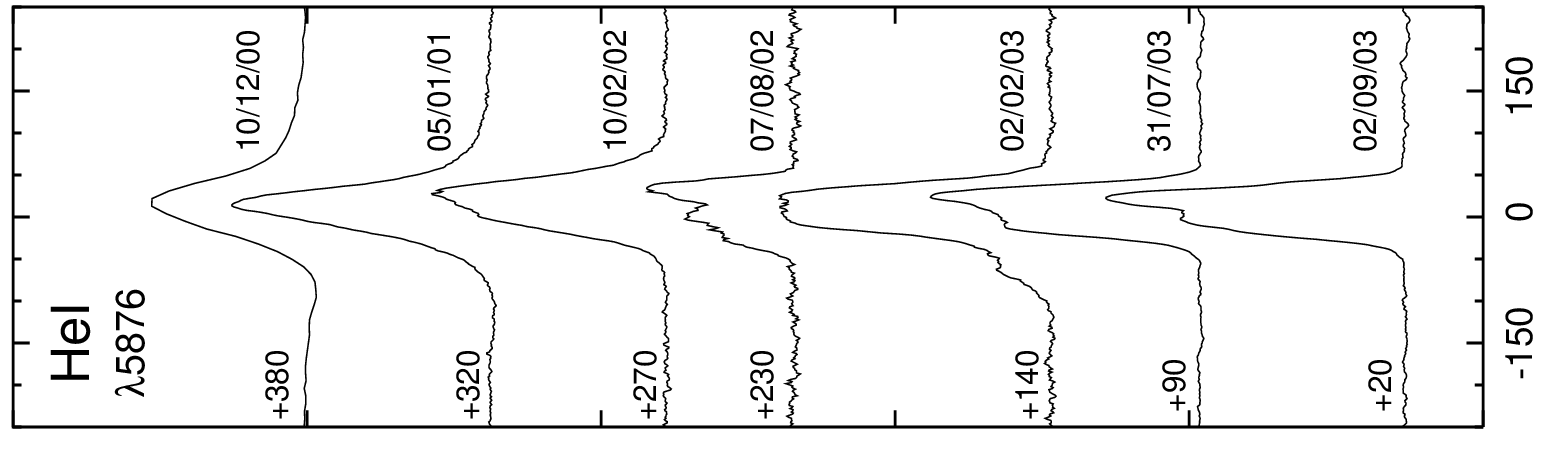}}

\vspace*{-60mm}
\hspace*{-37mm}
\resizebox{144mm}{!}{\includegraphics[angle=-90]{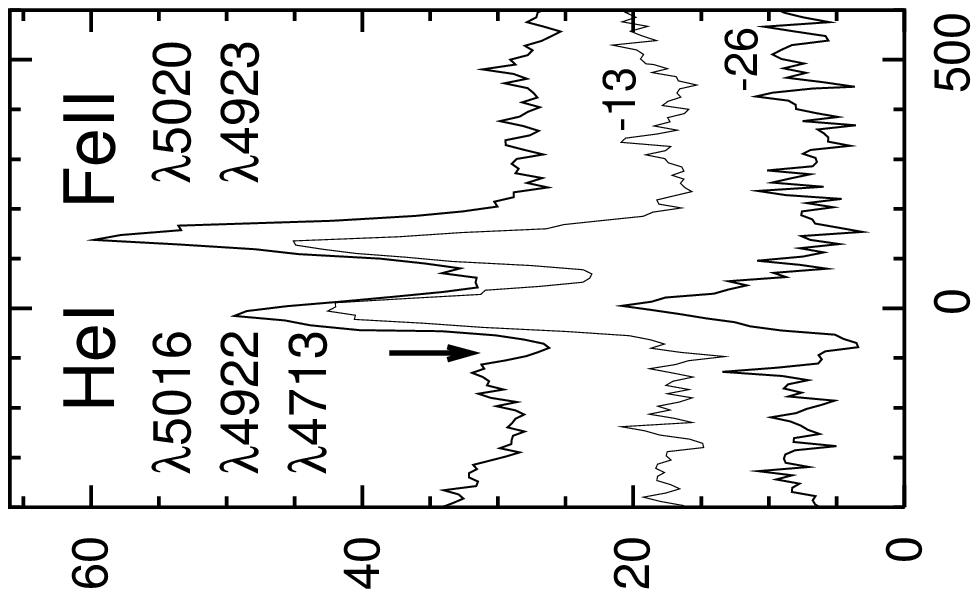}
                     \includegraphics[angle=-90]{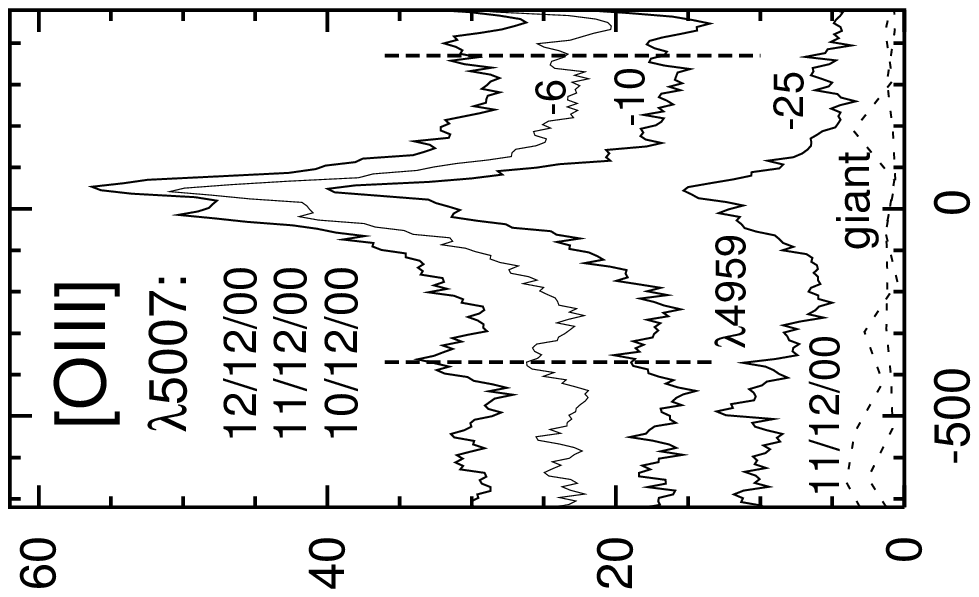}
                     \includegraphics[angle=-90]{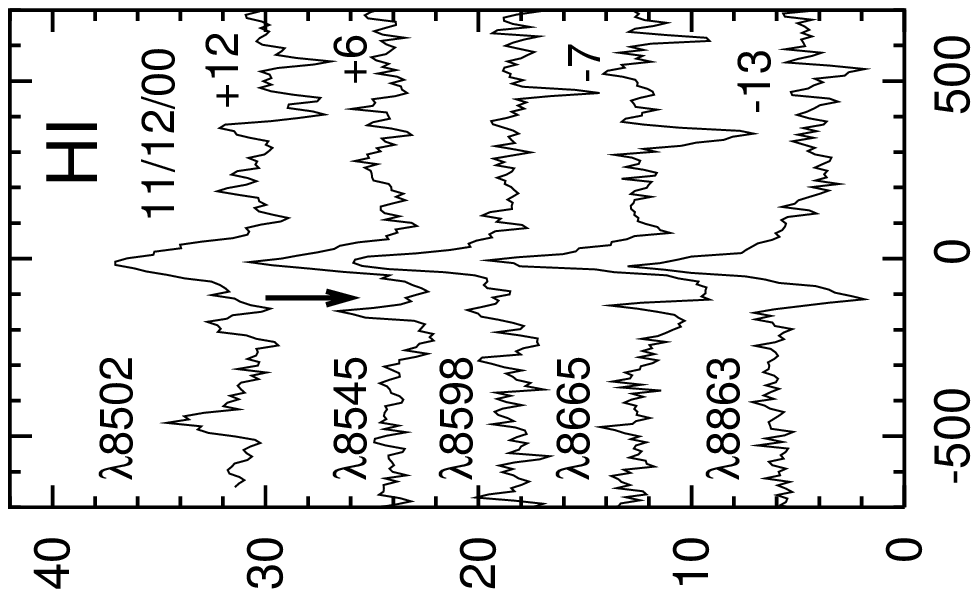}
                     \includegraphics[angle=-90]{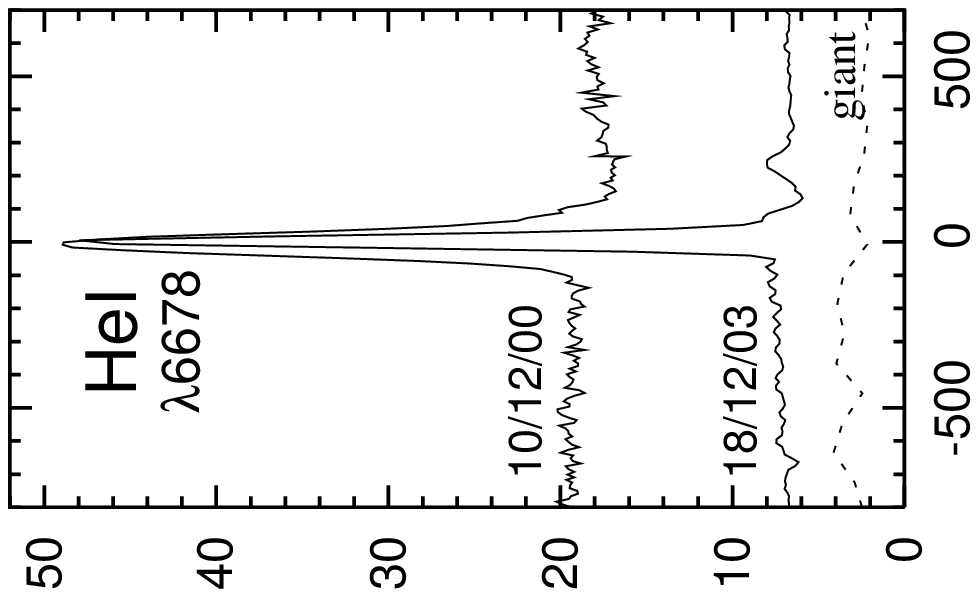}}
%o3orig.eps
\caption[]{
Evolution in selected lines, mainly from the optical maximum. 
Fluxes are in units of $10^{-13}$\ecsa\ and radial velocities 
in \kms. Small numbers in panels with more profiles represent 
their shift with respect to the level of the local continuum. 
The arrows denote absorption components. The right panel shows 
evolution in the \ion{He}{i}\,$\lambda$5876 line 
(y-tics are by 100 units). The continuum from the giant, as 
given by the SED (Fig.~7), is compared to clarify details in 
lines of H${\beta}$, [\ion{O}{iii}]\,$\lambda$5007 and 
\ion{He}{i}\,$\lambda$6678. 
          }
\end{center}
%\label{fig_1}
\end{figure*}
\section{Analysis}

\subsection{Photometric evolution}
%^^^^^^^^^^^^^^^^^^^^^^^^^^^^^^^^^

Bottom panels of Fig.~1 show the $UBVR$ LCs of Z\,And 
covering its recent outburst. The rise to maximum was 
characterized by three rapid increases in the star's 
brightness and two plateaus, best seen in the $U$ band. 
The initial rise between 2000 September 1st and 7th 
($\Delta U\dot = 1.9$, 
 $\Delta B\dot = 1.5$, 
 $\Delta V\dot = 1.1$)
was probably caused by an increase of the hot object luminosity. 
The $U-B$ index became more blue with respect to observations 
made prior to the outburst. Consequently, the profile of the 
optical continuum became more steep. 
Its nature -- nebular or stellar (or a combination of both) -- 
is important to understand the trigger mechanism of the outburst. 
The nebular nature of the optical light at very beginning of 
the outburst would require an increase in $T_{\rm h}$ to get 
the necessary amount of ionizing photons, while the stellar 
nature of the optical radiation would be associated with 
a creation of a shell surrounding the hot star implying thus 
a decrease in $T_{\rm h}$. Unfortunately, we do not have 
relevant spectroscopic observations from the initial stage 
of the activity to distinguish between these possibilities. 
Nonetheless a comparison of colour indices with those of 
a black body \citep[they are well above, see Fig.~1 of][]{sk03b} 
signals a contribution from other sources. 
The following short-term decline in the $U-B$ index at
constant $B-V$ and $V-R$ indices suggested a decrease of
the hot component temperature. Then during the 'totality' of 
the $U-B$ minimum ($\sim JD~2\,451\,800 - 2\,451\,865$) 
all the colour indices were practically constant indicating 
thus a constant temperature. Therefore the second increase 
in the star's brightness, which occurred at constant 
indices ($\sim JD~2\,451\,820$, Fig.~1), had to result from 
an increase in the luminosity at a relevant expansion of 
the shell. 
Finally, the third brightening from the mid-November to the 
maximum (i.e. the ascending branch of the $U-B$ minimum)
indicated an increase of the hot object temperature and its 
luminosity. Quantities of the corresponding colour indices 
were probably due to a composition of contributions from 
a nebula and stellar source. Our SED model from the maximum 
supports this interpretation (Sect.~3.3.2, Fig.~7). 
It is of interest to note that similar behaviour in colour 
indices was also observed at early stages of classical novae, 
prior to the maximum \citep[e.g. Fig.~3 of][]{ch+pr97}.
%
%
%==================================================|
%---- Fig. 3: FUSE feets of the OVI~1032 line  ----|
%==================================================|
% [p!th] [!ht]
%
\begin{figure}
\centering
\begin{center}
\resizebox{\hsize}{!}{\includegraphics[angle=-90]{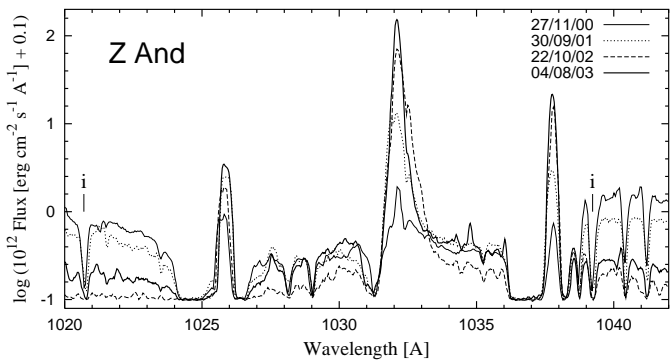}}
\caption[]{
Evolution in the \ion{O}{vi}\,$\lambda$1032 line (LiF1A spectra) 
in an appropriate logarithmic scale. Note that the P-Cygni 
profile with a strong absorption developed throughout the whole 
active phase. Positions of interstellar lines are denoted 
by {\bf i}. 
          }
\end{center}
%\label{fig_1}
\end{figure}
%
%
%==================================================|
%--- Fig. 4: H_alpha line series + Raman, FeVII ---|
%==================================================|
%
\begin{figure*}
\centering
\begin{center}
\resizebox{\hsize}{!}{\includegraphics[angle=-90]{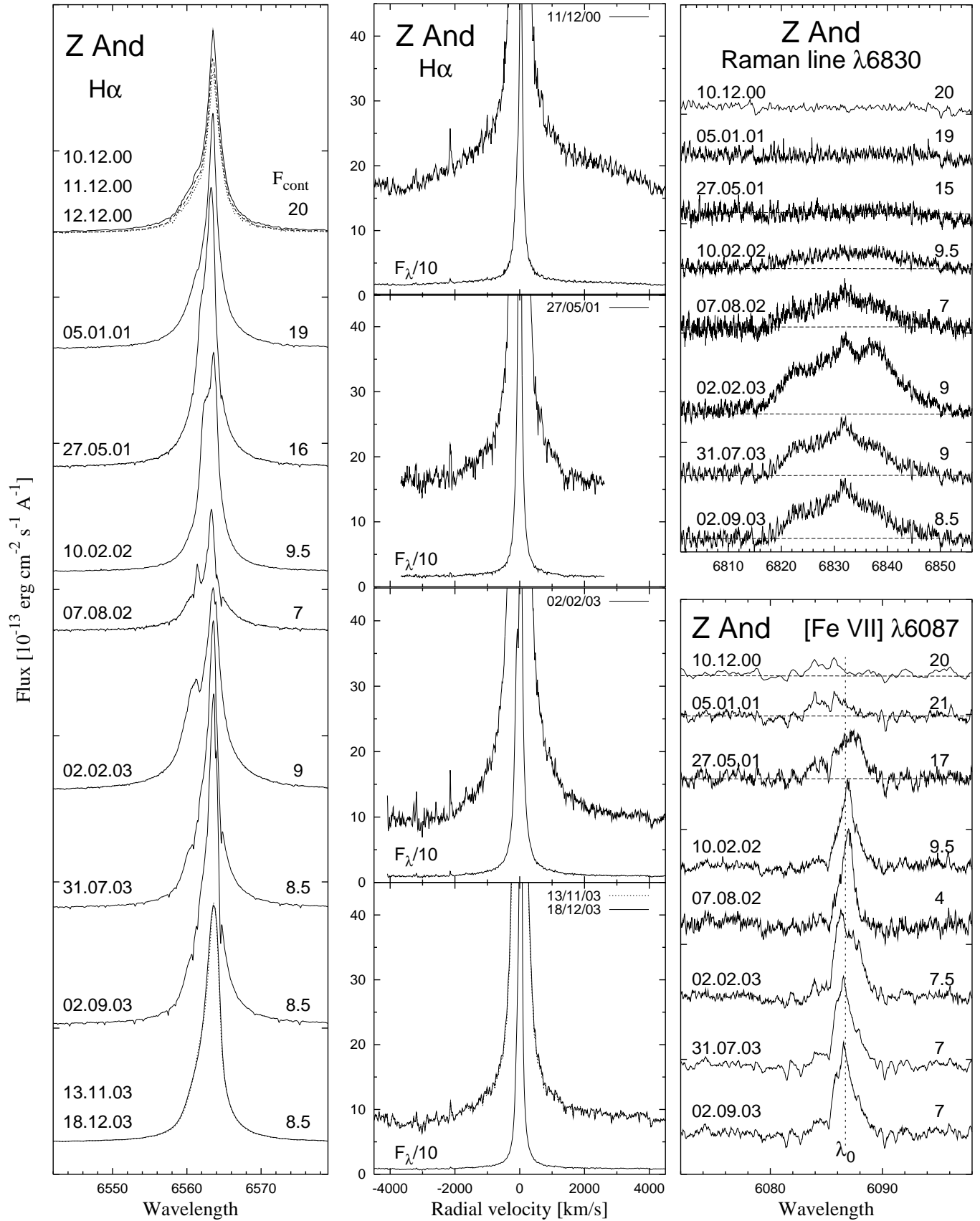}}
%\resizebox{15cm}{!}{\includegraphics[angle=-90]{sp_ev.eps}}
%
\caption[]{
Evolution in the H$\alpha$, Raman $\lambda$6825 and 
[Fe\,\V\I\I]\,6087 line profiles along the outburst, from 
the 2000-maximum to 2003-quiescence. Left panel shows 
profiles of the H$\alpha$ line core (y-tics are by 500 units), 
while the middle panels show the extended wings of 
the profiles in detail. On the right panels y-tics are 
by 20 units and the horizontal dashed lines represent 
the approximated continuum as described in Sect.~2.1.1. 
%$500\,10^{-13}$\ecsa
          }
\end{center}
%\label{fig_1}
\end{figure*}
%---------------------- odsek --------------

From 2001 Z\,And entered the decline phase during which both 
colour indices and the star's brightness were evolving 
toward quantities of a quiescent phase at the end 
of 2003 (Fig.~1). In the summer of 2002 the gradual decline 
was interrupted by the eclipse effect. This event revealed 
a high orbital inclination and the profile of the minimum 
suggested a disk-like structure for the eclipsed object 
\cite[see][in detail]{sk03a}. 
In addition, a transient $\sim$\,0.6\,mag rebrightening 
lasting from the autumn 2002 to the spring 2003 was indicated 
in the LC (see Sect.~3.6 in more detail). 

Our reconstructed SEDs, we present in the section 3.3, should aid 
to understand better the photometric behaviour we discussed 
qualitatively above. 
%
%==========================================|
%--- Fig. 5: FUSE 30/09/01 + 04/08/03  ----|
%==========================================|
%
\begin{figure}
\centering
\begin{center}
\resizebox{\hsize}{!}{\includegraphics[angle=-90]{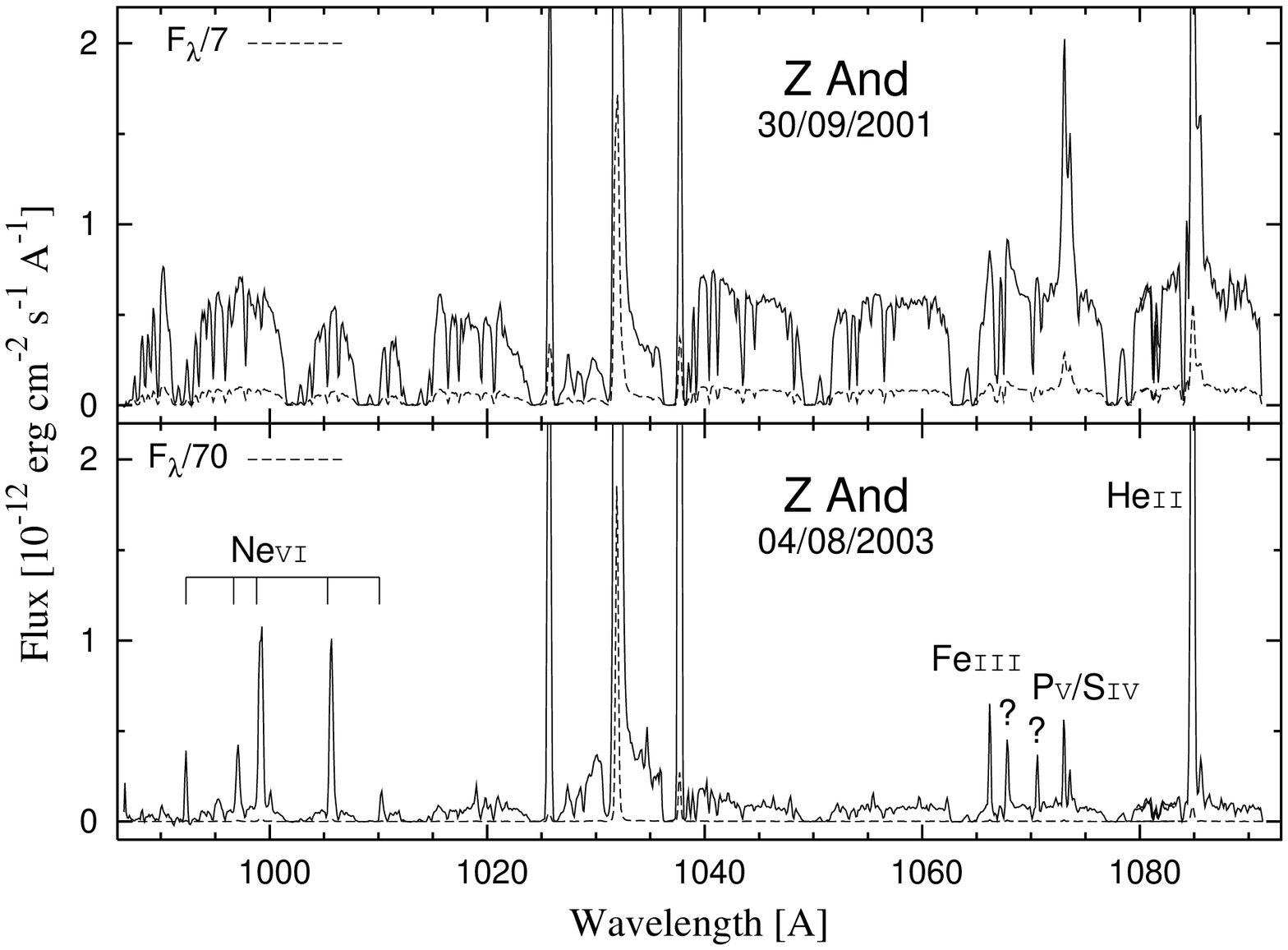}}
\caption[]{
Observed LiF1A spectra from 30/09/01 and 04/08/03. The level 
of the continuum significantly decreased while fluxes of 
the \ion{O}{vi} doublet increased by a factor of $\sim$\,10
on 04/08/03. Note also the appearance of the \ion{Ne}{vi} lines. 
          }
\end{center}
%\label{fig_1}
\end{figure}

\subsection{Spectroscopic evolution}
%^^^^^^^^^^^^^^^^^^^^^^^^^^^^^^^^^^^

Figures~2$\div$6 show main characteristics of our spectroscopic 
observations along the outburst, from its maximum 
in 2000 December to its end in 2003. 
We comment on them as follows: 

 (i) 
The P-Cygni profiles of \ion{H}{i} lines of the Paschen 
series and \ion{He}{i} lines were present mainly at initial 
stages of the outburst, prior to and at the maximum. 
%
%In the optical we measured them in the \ion{He}{i} lines 
%\,$\lambda$4713, $\lambda$4922, $\lambda$5016, 
%$\lambda$5876\,\AA\ 
%\ion{H}{i} lines of the Paschen series. 
%
The main absorption component was located at about $-90$\kms. 
However, complex absorption features to $\sim-300$\kms\ could 
be recognized for $\lambda$8502, $\lambda$8545 and $\lambda$8665. 
%
%A complex absorption/emission structure of the Paschen series 
%of hydrogen lines can be seen on the violet side from 
%the reference wavelength. For $\lambda$8502, $\lambda$8545 
%and $\lambda$8665, absorption features can be recognized 
%to $\sim-300$\kms. 
%In the H$\beta$ profile we identified 
%an absorption at $\sim-600$\kms. 
%
In the far-UV \cite{sok+05} detected P-Cygni profiles 
in \ion{C}{iii} and \ion{P}{v} lines (Fig.~6 here) 
with terminal velocities of $\approx-200$\kms\ on the 
first $FUSE$ observations. They also demonstrated evolution 
from absorption to emission profile of these lines as the 
outburst progressed to its maximum. On the other hand, 
the resonance line \ion{O}{vi}\,$\lambda$1032 shows a strong 
absorption in the P-Cygni line profile throughout 
the whole active phase (Fig.~3). This suggests its origin 
in the stellar wind of the hot star. During quiescence 
the absorption component was not present in the profile 
\citep{schmid+99}. In Sect.~3.4 we examine this line in detail. 

  (ii)
The H$\beta$ wings extended to about 1000\kms\ or more. 
The main emission core showed an asymmetrical profile with 
the peak placed at about $+30$\kms\ (the systemic velocity is 
$-1.8$\kms). The asymmetry can result from superposition of 
a violet shifted absorption component created in the neutral 
gas at the front of the expanding material. 
The H$\alpha$ profile exhibited similar behaviour. However, 
its wings extended to more than $\pm 2\,000$\kms\ and the asymmetry 
was more pronounced. At/after the spectroscopic conjunction 
the absorption component created a cut-out on the violet 
side of the wing at about $-40$\kms\ (07/08/02) and 
$-50$\kms\ (02/02/03). 
During the low-amplitude brightening (2002/03) we measured 
an additional extension of the H$\alpha$ wings on our spectrum 
from 02/02/03. 
%
%According to properties of the Ly$\beta$ line as observed by $FUSE$ 
%
We ascribe their origin to the stellar wind from the hot star 
(Sect.~3.5). 
%
%rather than to the Raman scattered Ly$\beta$ photons. 
%We examine this case in Sect.~3.5. 

 (iii) 
At the maximum, the main emission core of the 
\ion{He}{ii}\,$\lambda$4686 was shifted by about 
$-75$\kms\ at the peak and by about $-105$\kms\ at its 
bottom with respect to the reference wavelength. 
The [\ion{Fe}{vii}]\,$\lambda$6087 line displayed similar 
characteristics. Its  profile was very low and broad 
with the main emission located around $-80$\kms\ (Fig.~4). 
On the transition from the optical maximum to quiescence 
these lines became stronger being located at/around their 
reference wavelengths (see Fig.~4 for [\ion{Fe}{vii}] and 
Tomov et al.~2005 for the \ion{He}{ii} line). 
The evolution in these lines is discussed in Sect.~4. 
%
%results from the hot object 
%structure and its changes along the outburst (Sect.~4). 
%
%
%==========================================|
%----- Fig. 6: FUSE at maximum: n_H  ------|
%==========================================|
%
\begin{figure*}
\centering
\begin{center}
\resizebox{15cm}{!}{\includegraphics[angle=-90]{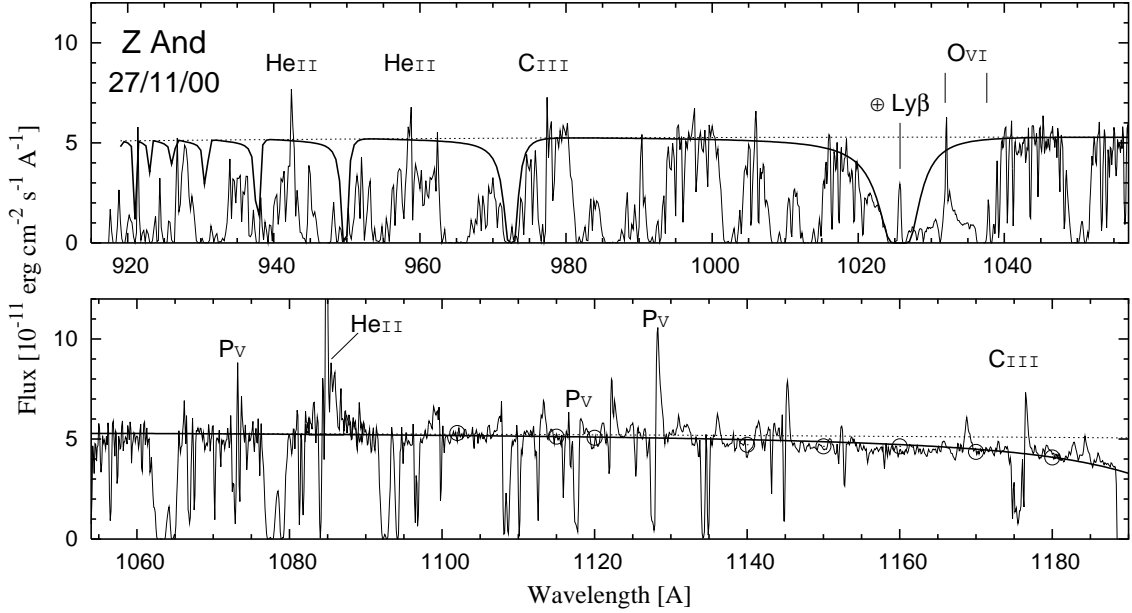}}
\caption[]{
Dereddened $FUSE$ spectra of Z\,And from the maximum of 
the optical brightness. It is composed from 
SiC2A (917$\div$1005\,\AA), 
LiF1A (988$\div$1082\,\AA) and 
LiF2A (1087$\div$1181\,\AA) channel spectra. 
Spectra from SiC1A (1080$\div$1090\,\AA) and LiF1B 
(1080$\div$1188\,\AA) have been used to complete the wavelength 
coverage. The solid sick line represents the Rayleigh attenuated 
($N_{\rm H} = 7\,10^{21}$\cmd) blackbody radiation from 
the hot stellar source at 28\,000\,K (Sect.~3.3.2), while the 
dotted line is not attenuated. Most of the absorption features 
are due to H$_{2}$ in the interstellar medium \citep[cf.][]{cem}. 
Note the P-Cygni type of profiles for \ion{C}{iii}, \ion{P}{v} 
lines and the broad emission in the \ion{He}{ii}\,$\lambda$1084 
profile. 
          }
\end{center}
%\label{fig_1}
\end{figure*}

  (iv) % - HeI 5876 + 6678 + (7065) lines:
\ion{He}{i} $\lambda$5876, $\lambda$6678 and $\lambda$7065 
emission lines can be used to diagnose the symbiotic nebula 
\citep{proga+94}. We observed all these lines just at the 
maximum, but later, along the investigated active phase, 
only $\lambda$5876 and $\lambda$6678 were measured. 
At the beginning of the outburst they developed a P-Cygni type 
of profiles. From 2002 August absorptions affected the whole 
profile, mainly that of the $\lambda$5876 line (Fig.~2). 
A narrow emission core without any wings and with absorption 
features at the top of the profile were observed. 
In addition, ratios of their fluxes differ significantly 
from those arising from net recombination. For example, 
we measured the ratio $\Lambda = F(5876)/F(6678)\sim 1.2\div1.5$, 
while theoretical intensities from low-density plasma predict 
$\Lambda \sim 3.6$ \citep[e.g.][]{smits}. 
This difference has been ascribed to collisions and optical 
depth due to the metastability of the $2\,^{3}$S level, 
which affect the \ion{He}{i} level populations in a high 
density plasma \citep[e.g.][]{clegg,schmid89a,proga+94,kf}. 
Therefore we compared our flux ratios, $F(6678)/F(5876)$ and 
$F(7065)/F(5876)$ with those modeled by \cite{proga+94} for 
the case of high density nebulae in symbiotic stars 
(Appendix~A, Fig.~A.1). This suggests that the \ion{He}{+} 
region is very compact with an average density 
$N_{\rm e} \approx 10^{12}$\cmt, radiating at 
$T_{\rm e} \approx 20\,000$\,K. Emissivity of such the 
plasma in the \ion{He}{i}\,$\lambda$5876 line corresponds 
%
%[$\varepsilon_{5876} \dot = 1\,10^{-26}$\eps\ 
%\citep[see Fig.~2 of][]{proga+94} and fluxes in Table~A.1] 
%
to an effective radius, 
   $R_{\rm eff}(\ion{He}{+}) \approx 5\,R_{\sun}$, 
which represents the radius of a sphere with constant density 
(see Appendix~A in more detail). 

  (v)  
Profile of the nebular [\ion{O}{iii}]\,$\lambda$5007 line 
showed asymmetrically extended core/wings with terminal 
velocities of about -300 and +250\kms. A sequence of profiles 
from 2000 December 10, 11 and 12th, showed a creation of 
an extra emission component on the violet side of the main 
core. 
In addition, weak emission components at about $\pm 380$\kms\ 
accompanied the main core during the maximum. Their reality 
is supported by the S/N ratio of about 10 at these wavelengths 
and a relatively low level of the continuum from the giant 
(cf. Figs.~2 and 7). 
If one associate these components to polar blobs, as was 
interpreted for the classical Nova V\,1974 by \cite{ch+97}, 
the space velocity of the ejecta would be larger than 
$\sim$1\,600\kms\ for $i > 76^{\circ}$ \citep{sk03a}. 

  (vi) 
Evolution in the Raman scattered \ion{O}{vi} line at 
$\lambda$6825 provides information on the ionization structure 
around the active object. This emission was absent on our 
spectra from the maximum and after it. It appeared first on 
the 10/02/02 spectrum as a very faint emission. During 2003 
it became strong and broad (FWZI\,$\approx$\,25$\div$30\,\AA) 
with 1$\div$3 knots at the top of its profile (Fig.~4). 
\cite{schmid89b} suggested that this emission feature is due 
to Raman scattering of the \ion{O}{vi}\,$\lambda$1032 
resonance line by atoms of neutral hydrogen. In Sect.~3.4 
we investigate this process in detail. 

  (vii) 
Transition to quiescence was characterized mainly by a significant 
increase in the hot object temperature (Sect.~3.3.3). 
As a result fluxes of highly ionized elements 
increased and new appeared (e.g. \ion{Ne}{vi} lines, Fig.~5). 
At the end of 2003 a constant brightness near to the values 
from quiescence and a very stable H$\alpha$ profiles 
(13/11/2003 and 18/12/2003) suggested the Z\,And active phase 
to be over. 
However, a strong wind from the hot star was still indicated 
by the P-Cygni profiles of the \ion{O}{vi} line (Fig.~3), and 
in 2004 August we detected a new eruption in the LC (Fig.~1). 

\subsection{Spectral energy distribution}
%^^^^^^^^^^^^^^^^^^^^^^^^^^^^^^^^^^^^^

Here we aim to reconstruct the SED prior to the maximum, at 
the maximum and during the post-outburst stage to understand 
better the observed photometric and spectroscopic evolution 
throughout the outburst. 

\subsubsection{Model of the composite spectrum}
%^^^^^^^^^^^^^^^^^^^^^^^^^^^^^^^^^^^^^^^^^^^

According to a simple photoionization model for symbiotic stars 
proposed by \cite{stb} (hereafter STB), the observed flux in 
the continuum, $F(\lambda)$, can be expressed as 
%
% -------------------- Eq. (1) ------------------------
%
\begin{equation}
  F(\lambda) =  F_{\rm g}(\lambda) + F_{\rm h}(\lambda) +
                F_{\rm N}(\lambda),
\end{equation}
%------------------------------------------------------
%
where $F_{\rm N}(\lambda)$ is the nebular component of radiation 
and $F_{\rm g}(\lambda)$ and $F_{\rm h}(\lambda)$ are stellar 
components from the hot star and the cool giant, respectively. 
We approximated radiation from the giant by a synthetic 
spectrum, $F^{\rm synth.}_{\lambda}(T_{\rm eff})$ 
\citep[models from][]{h+99}, the hot stellar continuum by 
a black body at the temperature $T_{\rm h}$ including 
the effect of Rayleigh scattering and the nebular radiation 
in the continuum by processes of recombination and thermal 
bremsstrahlung in the hydrogen plasma. Then Eq.~(1) takes 
the form \citep[][Eq.~13]{sk05}
%
%-------------------- Eq. (2) ----------------------
\begin{equation}
 F_{\lambda} = F^{\rm synth.}_{\lambda}(T_{\rm eff}) +
\theta_{\rm h}^2 \pi B_{\lambda}(T_{\rm h})
       e^{-n_{\rm H}\sigma_{\lambda}^{\rm R}} +
\frac{EM}{4\pi d^2}
      \varepsilon_{\lambda}({\rm H},T_{\rm e}),
\end{equation}
%--------------------------------------------------------------
%
where $\theta_{\rm h} = R_{\rm h}^{\rm eff}/d$ is the angular
radius of the hot stellar source given by its effective 
radius $R_{\rm h}^{\rm eff}$ and the distance $d$. Parameters 
determining the Rayleigh attenuation are the column density 
of H atoms, $n_{\rm H}$ [cm$^{-2}$], and the Rayleigh 
scattering cross-section for atomic hydrogen, 
$\sigma_{\lambda}^{\rm R}$ [cm$^{2}$]. Nebular radiation 
is in our approximation defined by the emission measure 
$EM$ and the volume emission coefficient for hydrogen 
$\varepsilon_{\lambda}({\rm H},T_{\rm e})$ 
[erg\,cm$^{3}$\,s$^{-1}$\,\AA$^{-1}$], which depends on 
the electron temperature $T_{\rm e}$. 
The method of disentangling the composite spectra 
of symbiotic stars was recently described by \cite{sk05}. 

To reconstruct the continuum of the hot object we used only 
the flux-points of the $UBVR$ photometry and the far-UV
continuum from $\lambda$1188 to $\lambda$1100\,\AA. 
With the aid of the $FUSE$ spectra we determined 
the column density of the atomic hydrogen by fitting the 
Rayleigh attenuated hot stellar radiation to the observed 
continuum blueward Ly$\beta$ and Ly$\alpha$ lines (Fig.~6). 
This reduced free parameters in the model to $\theta_{\rm h}$, 
$T_{\rm h}$, $T_{\rm e}$ and $EM$. 
As there was no ultraviolet spectrum available from the 
2000-03 outburst, determination of $T_{\rm e}$ is more 
uncertain. 
To satisfy the condition for an appropriate fit 
\citep[][Sect.~3.3]{sk05} we estimated the uncertainty in 
$T_{\rm h}$ within the range of 500$\div$1\,000\,K, 
while $T_{\rm e}$ can run between about 30\,000 and 40\,000 K. 

\subsubsection{SED at the plateau stage and the maximum}
%^^^^^^^^^^^^^^^^^^^^^^^^^^^^^^^^^^^^^^^^^^^^^^^^^^^^

To reconstruct the SED during the time of the last plateau 
in the LC we used the optical spectrum from 06/11/00 and 
the $FUSE$ observation from 16/11/00. For the maximum, 
the corresponding data are from 11/12/00 and 15/12/00, 
respectively. 
A comparison of our solutions with observations is plotted 
in Fig.~7 and the corresponding parameters are in Table~2. 

The change in the SED from the plateau at $U \sim 9.0$ 
to the maximum ($U \sim 8.3$) was due to an increase in 
both the hot stellar and the nebular emission. 
These components of radiation contributed more to the $U$ band 
than to the $BVR$ region. Therefore we observed a larger 
increase of the star's brightness in $U$ 
($\Delta U\sim 0.7,~\Delta B\sim 0.4,~\Delta V\sim 0.3, 
~\Delta R\sim 0.3$\,mag), i.e. the $U-B$ index became more 
blue. The temperature of the hot stellar source, $T_{\rm h}$, 
increased from about 26\,000\,K to about 28\,000\,K at 
a constant effective radius $R_{\rm h}^{\rm eff}$ (Table~2). 
Accordingly, its luminosity increased from 3\,800 to 
5\,300\,$(d/1.5\,{\rm kpc})^{2}\,L_{\sun}$ during this 
transition. However, the emission measure also increased; 
significantly by a factor of 2.2. 

The {\em observed} hot stellar radiation is not capable of 
producing the nebular emission (too low $T_{\rm h}$). 
%
%\citep[the parameter $\delta = EM_{\rm obs}/EM_{\rm max} > 1$, 
%see Sect.~4.1 of][]{sk05}. 
%
This implies that the hot ionizing source {\em cannot} be seen 
directly by the outer observer, but {\em can} be 'seen' by 
the surrounding nebula, which converts its radiation into 
the nebular emission. 
This suggests a disk-like structure of the optically thick 
material, whose outer rim occults the central hot star in 
the direction of the observer 
\citep[also Sect.~5.3.5 of][]{sk05}. 
For comparison we show the SED from the previous 1986 optical 
maximum, which is of the same type. 
In such the case the temperature of the ionizing source, 
$T_{\rm h}^{\rm i.s.}$, can be determined only {\em indirectly} 
from the nebular radiation, which mirrors its radiative 
properties. For example, \cite{sok+05} determined 
$T_{\rm h}^{\rm i.s.}$ = 95\,000\,K from fluxes of \hb and 
\ion{He}{ii}\,4686 emission lines observed at the maximum. 
However, scaling the hot radiation from the unseen ionizing 
source to the observed, significantly cooler, pseudophotosphere 
yields an erroneous (too high) luminosity of the white dwarf. 
According to our SED model the total luminosity produced 
at/around the white dwarf during the maximum is 
$L_{\rm T} \ga L_{\rm h} + L_{\rm N} = 
                          6\,900\,(d/1.5\,{\rm kpc})^2\,L_{\sun}$ 
\citep[Table~2 here and Sect.~5.3.6 of][in more detail]{sk05}, 
but \cite{sok+05} derived 
$L_{\rm h} = 9\,800\,(d/1.2\,{\rm kpc})^2\,L_{\sun}$ for 
$E_{\rm B-V}$ = 0.27 
         ($= 21\,500\,(d/1.5\,{\rm kpc})^2\,L_{\sun}$ for 
$E_{\rm B-V}$ = 0.30) 
by scaling the 95\,000\,K radiation of the ionizing source 
to the observed FUV fluxes from the warm pseudophotosphere. 
Also the best model SED of the FUV and optical continuum 
with the fixed stellar component radiating at 
$T_{\rm h}^{\rm i.s.}$ = 95\,000\,K is not satisfactory 
($\chi \dot = 56.2$). For comparison, our models in mid panels 
of Fig.~7 have $\chi \dot = 8.5$ and $\chi \dot = 13.6$, 
respectively. 
%
%==========================================|
%---------- Fig. 7: Z And SED's  ----------|
%==========================================|
%
\begin{figure}[p!t]
\centering
\begin{center}
\resizebox{\hsize}{!}{\includegraphics[angle=-90]{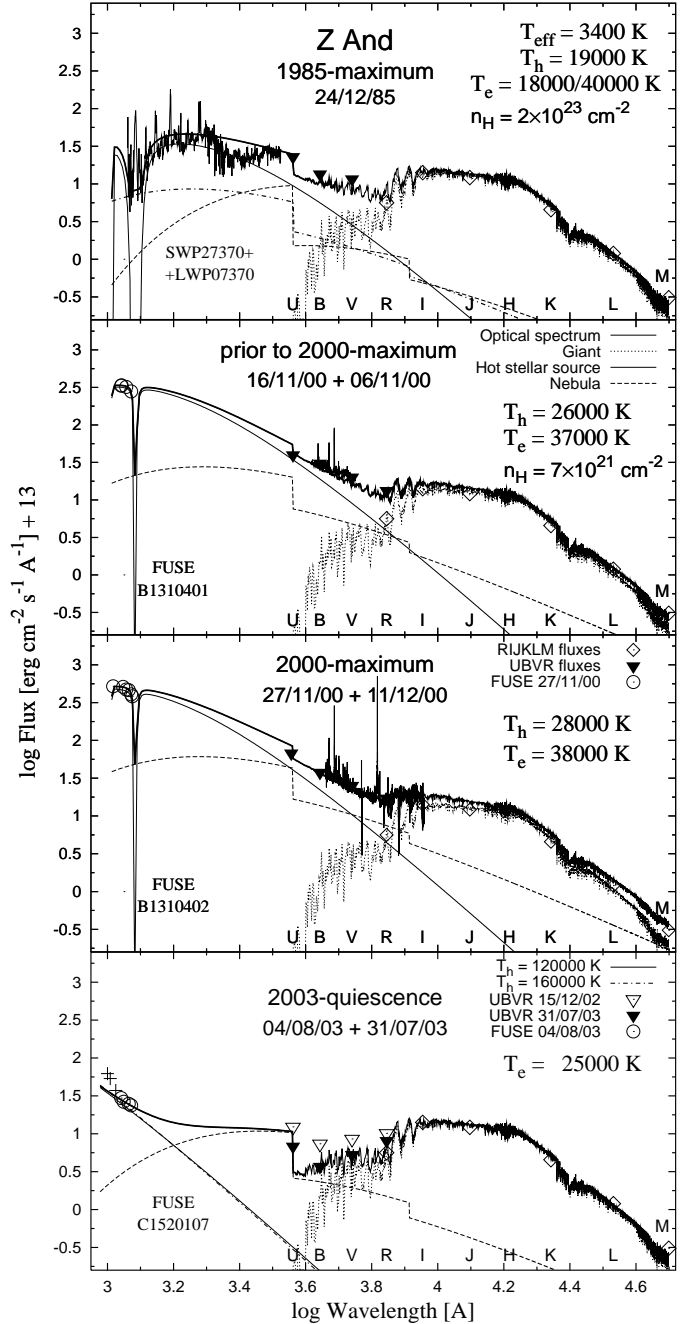}}
\caption[]{The UV/optical/IR SED during active phases of Z\,And. 
Top: At the 1985-maximum. 
Middle two panels: Prior to and at the maximum of the 2000 
         outburst (positions in the LC are shown in Fig.~1). 
Bottom: At the post-outburst stage in the summer of 2003. 
The resulting spectrum (solid thick lines) is given by 
superposition of nebular component(s) (dashed lines) 
and two stellar components, from the hot stellar source 
(solid thin lines) and the giant (dotted lines). 
$IR$ fluxes ($\Diamond$) are from quiescence and far-UV fluxes 
($\circ$, $+$, Sect.~2.2) are near-simultaneous to 
the optical observations. 
           }
\end{center}
%\label{fig_1}
\end{figure}
%
%
%==============================================================|
%-- Table 2: Parameters of SEDs: Z And 1985 + 2000-outburst ---|
%==============================================================|
%
\begin{table*}[p!t]
\begin{center}
\caption{Fitting and derived parameters from the SED during the maximum 
         of the 1985 and 2000 outburst.}
\begin{tabular}{rcccccccc}
\hline
\hline
Date~~~~~~~~~~~~~~&
$T_{\rm h}$       &
$\theta_{\rm h}$  &
$R_{\rm h}^{\rm eff}$&
$L_{\rm h}$       &
$\log(n_{\rm H})$ &
$T_{\rm e}$       &
$EM$              &
$L_{\rm N}$        \\
%$EM_{\rm HTN}$    &
%$L_{\rm HTN}$     \\
%
$FUSE$\,--~Optical~~~&
[$K$]             &
                  &
[$R_{\sun}$]      &
[$L_{\sun}$]      &
[cm$^{-2}$]       &
[$K$]             &
[cm$^{-3}$]       &
[$L_{\sun}$]       \\
%[cm$^{-3}$]       &
%[$L_{\sun}$]     \\
%
%-------  T_h,  O_h,  R_h,  L_h,  n_H,  T_e,  EM,  L(N)
%
\hline
          24/12/85& 19\,000& 3.4\,10$^{-11}$& 2.3&  ~~590& 23.34& 18\,000/40\,000& 
                    2.1\,10$^{60}$&  425     \\
16/11/00--06/11/00& 26\,000& 4.6\,10$^{-11}$& 3.1& 3\,800& 21.85& 37\,000        &
                    3.9\,10$^{60}$&  740     \\
27/11/00--11/12/00& 28\,000& 4.7\,10$^{-11}$& 3.1& 5\,300& 21.85& 38\,000        &
                    8.8\,10$^{60}$& 1\,600~~~\\
04/08/03--31/07/03&120\,000~~& 1.5\,10$^{-12}$& 0.10&1\,800&  -- & 25\,000&
                    1.3\,10$^{60}$&  280     \\
                  &160\,000~~& 1.3\,10$^{-12}$& 0.086&4\,400&  -- &-- " --&
                   -- " -- & -- " --         \\
%
%---------------------------------------------------------------------
%-------  T_h,  O_h,  R_h,  L_h,  n_H,  T_e,  EM,  L(N)
%
\hline
%\hline
\end{tabular}
\end{center}
%Parameters, $R_{\rm h}^{\rm eff}$, $L_{\rm h}$ and 
%$EM$ are scaled with $d$ = 1.5\,kpc for fluxes dereddened 
%      with $E_{\rm B-V}$ = 0.30. \\
\end{table*}

\subsubsection{SED at the post-outburst stage}
%^^^^^^^^^^^^^^^^^^^^^^^^^^^^^^^^^^^^^^^^^^

Here we selected the last observations made by the $FUSE$ 
on 04/08/03 and corresponding $UBVR$ photometric measurements. 
The slope of the $B$, $V$ and $R$ fluxes has an opposite trend 
to that from the maximum -- the fluxes follow rather the SED 
of the giant, which dominates the spectrum from the $B$ band 
to longer wavelengths (Fig.~7). 
The $U$ flux-point does not obey this trend. It is higher, 
the ratio $F_{\rm U}/F_{\rm B} \sim 1.78$ (i.e. $U - B$ = -1.16). 
This reflects a dominant contribution from the nebula in 
the near-UV/U. 
As a result the radiation from the hot star in the optical 
had to be small relatively to other sources here. 
This suggests a high temperature, $T_{\rm h} \ga 100\,000$\,K. 
We adopted $T_{\rm h} = 120\,000$\,K to satisfy basic 
{\em photo}-ionization model of symbiotic stars (STB). 
At this temperature the hot star luminosity is just capable 
of producing the observed $EM$. From this point of view 
$T_{\rm h} \ge 120\,000$\,K. 
Finally, we complemented contributions from the giant and 
the hot stellar source by a nebular contribution with 
  $T_{\rm e}$ = 25\,000\,+5\,000/-\,7\,000\,K 
and
  $EM = 1.3\,10^{60}$\cmt. 
Large uncertainties in $T_{\rm e}$ are given by the absence 
of ultraviolet data. Model parameters are in Table~2. 

\cite{sok+05} determined 
  $T_{\rm h} = (160\,000 \pm 35\,000)$\,K 
from emission line fluxes and 
  $R_{\rm h}^{\rm eff} = (0.06 \pm 0.01) \times 1.47
                       = (0.088 \pm 0.015)\,R_{\sun}$, 
  $L_{\rm h} = (2000 \pm 500) \times 2.20
             = (4400 \pm 1100)\,L_{\sun}$ 
from the $FUV$ continuum fluxes. The factors 1.47 and 2.20 convert 
their original quantities calculated for $E_{\rm B-V}$ = 0.27 
and $d$ = 1.2\,kpc to those corresponding to $E_{\rm B-V}$ = 0.30 
and $d$ = 1.5\,kpc we adopted in this paper. 
In this case physical parameters derived from the SED (Table~2) 
are the same as those given by optical line fluxes in combination 
with the $FUV$ continuum fluxes. 
This reflects an optically thin regime in 2003 -- in contrast 
to the situation from the optical maximum. 
%
%Compared are also $UBVR$ flux-points from the maximum of the 
%2002/03 brightening. These fluxes suggest the same type of 
%the SED as in the mid-2003, but with a stronger contribution 
%from the nebula. 

\subsubsection{SED as an indicator of the mass outflow}
%^^^^^^^^^^^^^^^^^^^^^^^^^^^^^^^^^^^^^^^^^^^^^^^^^^^^^^

The total emission measure of the nebular continuum at the 
maximum, $EM \sim 6\,10^{60}$\cmt\ (Table~2), was a factor 
of 5$\div$10 larger than in quiescence \citep[][]{nv89,sk05} 
and a factor of 3 larger than at the maximum of the 1985 
eruption \citep{fc+95}. 
During quiescent phases the nebula originates mostly from 
ionization of the giant's wind. It is characterized by a large 
parameter $X > 10$ in the STB model \citep[e.g.][]{fc+88,sk01}, 
which corresponds to a very open nebula (i.e. in major part 
a particle bounded). 
Therefore the significant increase of the nebular emission 
during the maxima of the outbursts can result from a supplement 
of new emitters into the particle bounded nebula, accompanied 
probably by an increase in the flux of ionizing photons 
from the hot source. The former can be caused by an increase 
in the mass-loss rate from the hot object. 
The broad H$\alpha$ wings (Fig.~4, Sect.~3.5) suggest the 
mass-outflow to be in the form of a stellar wind. 
\cite{sk06} estimated the corresponding mass-loss 
rate at the maximum as $\sim 1\div 2\,10^{-6}$\myr, 
which is a factor of $\sim$\,3 larger than that from the 
giant \citep[see Table~3 in][]{sk05}. 
The presence of a rather high-temperature nebula 
($T_{\rm e} = 30\,000\div40\,000$\,K, Table~2) requires a large 
contribution from collisions \citep{gur}, which is consistent 
with the strong mass-outflow. Also \cite{sok+05} ascribed 
the increased $X$-ray emission at the optical maximum to 
the shock-heated plasma as a consequence of the mass ejection 
from the white dwarf into the dense symbiotic nebula. 
Also the transient increase of the nebular emission during 
the low-amplitude 2002/03 rebrightening (bottom panel of 
Fig.~7) was also connected with a release of new emitters 
into the particle bounded nebula (see Sect.~3.6).
Contrarily, during the plateau stages the SED did not change 
considerably as suggested by the stability of colour indices 
(Sect.~3.1). This constrains the hot object to be stable in the 
luminosity, temperature and also the mass outflow, because 
more emitters would give rise a surplus of the emission 
measure and thus the star's brightness. 

\subsection{Direct and Raman scattered \ion{O}{vi}\,$\lambda$1032 line}
%^^^^^^^^^^^^^^^^^^^^^^^^^^^^^^^^^^^^^^^^^^^^^^^^^^^^^^^^^^^^^^^^^^^

The Raman scattering process and its possible applications 
in astrophysics was introduced by \cite{nsv89}. 
In this process a photon excites an atom from its ground state 
to an intermediate state being immediately stabilized by 
a transition to a true bound state by emitting a photon of 
a different frequency. 
\cite{schmid89b} identified the $\lambda$6830 and $\lambda$7088 
emission bands observed in many symbiotic stars as to be due 
to Raman scattering of the \ion{O}{vi} resonance doublet 
$\lambda$1032, $\lambda$1038 by neutral hydrogen. 
Later on this process has been investigated by many authors 
\citep[e.g.][]{schmid96,hh96,schmid+99,bes00}. 
The efficiency of this scattering process and profiles of both 
the direct and the Raman-scattered \ion{O}{vi} lines together 
with polarimetric measurements provide information on 
the scattering geometry in symbiotic systems \citep{schmid98}. 
Previous studies referred exclusively to quiescent phases 
of symbiotic stars. For Z\,And, \cite{bes98} revealed firstly
a strong evidence for the Raman effect and determined its 
efficiency to $\sim$\,7\,\%\ for $\lambda$1032 on the basis 
of the $HUT$ observations. \cite{schmid+99} derived 
the efficiency of this conversion to 4$\div$6\%\ and found 
that the Raman $\lambda$6825 profile in Z\,And is by 30\%\ 
broader than the original \ion{O}{vi}\,$\lambda$1032 line 
on the violet side of the profile. They used observations 
from the $ORFEUS$ mission. 

Here we analyze our series of Raman $\lambda$6830 lines 
obtained along the major Z\,And outburst (Fig.~4) together 
with the archival $FUSE$ spectra covering the same period. 

\subsubsection{Raman scattering efficiency}
%^^^^^^^^^^^^^^^^^^^^^^^^^^^^^^^^^^^^^^^

The scattering efficiency for the Raman process is defined as 
the photon ratio, $\eta(\ion{O}{vi}) = N_{\rm Ram}/N_{\ion{O}{vi}}$, 
between the Raman scattered and the initial \ion{O}{vi} line 
component, i.e. the fraction of emitted \ion{O}{vi} photons 
converted to Raman photons. 
As we measured only the Raman $\lambda$6825 line we investigate 
here the conversion $\lambda1032 \rightarrow \lambda6825$. 
In the simplest case the efficiency of this transition can 
be written as 
\begin{equation}
  \eta(\ion{O}{vi}) = 
            6.614\,\frac{F_{\rm Ram}}{F^{\rm obs}_{\ion{O}{vi}}},
\end{equation}
where the factor 6.614 = $\lambda1031.92/\lambda6825.44$. 
In the real case $\eta$ depends on the geometry of \ion{H}{i} 
atoms within the binary and the binary position with respect 
to the observer \citep{schmid96}. Thus the relation (3) 
represents just an empirical measure of the line photon ratio 
we call efficiency. 
Table~3 and Fig.~8 summarize the results. 
As the $FUSE$ and our optical observations were not simultaneous, 
we interpolated $FUSE$ fluxes to dates of optical observations 
and vice versa. Below we comment on the variation in $\eta$ 
along the outburst as follows:
%
%==================================================|
%-- Table 3: Measured fluxes and Raman efficiency -|
%==================================================|
%
\begin{table}[p!t]
\caption[]{Dereddened fluxes for the \ion{O}{vi}\,$\lambda$1032 
           line and its Raman scattered counterpart at $\lambda$6825 
           in units of $10^{-12}$\ecs. The efficiency, $\eta$, was 
           derived according to the relation (3).}
\begin{center}
\begin{tabular}{cccccc}
\hline
\hline
 Date & 
 Julian date & 
 Phase & 
 $F_{\ion{O}{vi}}$ & 
 $F_{\rm Ram}$ & 
 $\eta$  \\
      &  
 2\,4...    &
            & 
            &
            &
 [\%]    \\
\hline
%  Date       JD     Phase   F(1032)     6825      eta 
%--------------------------------------------------------------------
 16/11/00 & 51865.0 &0.16 &   50.6\,~&     --~   & --   \\  
 27/11/00 & 51876.0 &0.18 &   44.6\,~&     --~   & --   \\ 
 10/12/00 & 51889.4 &0.19 &   45.4\,i&  $<2^{\dagger}$  ~&$<29$ \\
 11/12/00 & 51890.4 &0.19 &   45.5\,i&  $<2^{\dagger}$  ~&$<29$ \\
 12/12/00 & 51891.3 &0.20 &   45.6\,i&  $<2^{\dagger}$  ~&$<29$ \\
 15/12/00 & 51894.0 &0.20 &   45.8  ~&     --~   & --   \\
 05/01/01 & 51914.9 &0.23 &   74.5\,i&  $<2^{\dagger}$  ~&$<18$ \\
 27/05/01 & 52057.3 &0.42 &  270.\,i~~~&$<2^{\dagger}$  ~& $<5$ \\
 22/07/01 & 52113.3 &0.49 &  347.~~~~~&    --~   & --   \\
 30/09/01 & 52183.3 &0.58 &  308.~~~~~&    --~   & --   \\
 10/02/02 & 52315.9 &0.76 &  972.\,i~~~&    4.31 ~& 2.9 \\
 05/07/02 & 52461.5 &0.95 & 1700.~~~~~~&   8.93\,i & 3.5\\
 07/08/02 & 52494.3 &0.99 & 1520.\,i~~~~&   9.97 ~& 4.3 \\
 22/10/02 & 52570.5 &0.09 & 1103.~~~~~~&  16.2\,i~& 9.7 \\
 02/02/03 & 52673.5 &0.23 & 1303.\,i~~~~&  24.7\,~~&12.5~~\\
 31/07/03 & 52852.5 &0.47 & 1650.\,i~~~~&  13.7\,~~& 5.5 \\
 04/08/03 & 52856.8 &0.47 & 1658.~~~~~~&  13.8\,i~& 5.5  \\
 02/09/03 & 52885.5 &0.51 & 1714.\,e~~~~&  14.4\,~~& 5.6 \\
\hline
\end{tabular}
\end{center}
  i/e -- interpolated/extrapolated value, \\
  ${\dagger}$ -- no detection, upper limits only
\end{table}

  (i) 
At the initial stages (2000-01) the not-measurable 
Raman's band could be a consequence of a very low 
\ion{O}{vi}\,$\lambda$1032 flux. Assuming the conversion 
efficiency of a few percent, the incident flux of 
$45\times10^{-12}$\ecs\ (Table~3) produces a few times 
$10^{-13}$\ecs\ at $\lambda$6825, which implies the height of 
the Raman feature above the local continuum only of a few 
times $10^{-14}$\ecs, which is under the detection. 
On the other hand, the signal to noise ratio S/N$\sim$20 
around the Raman line suggests a limiting flux of 
$\sim 1\times10^{-13}$\ecsa\ to recognize a real detail in 
the spectrum. Then, assuming the 
Raman line to be flat and broad as $\sim$20\,\AA, its flux 
we cannot detect is $\la\,2\times10^{-12}$\ecs. Corresponding 
upper limits for efficiencies $\eta$ are relatively large, 
because of low initial \ion{O}{vi} fluxes at the optical 
maximum (Table~3, Eq.~(3), Appendix~B). 

  (ii) 
In the period of 2002, prior to the spectroscopic conjunction, 
the emergence of the Raman features and the enhancement of 
the \ion{O}{vi} fluxes (Table~3, Fig.~8) require a relevant 
extension of the scattering region to get the measured 
efficiencies of 3$\div$4\,\%. Note that the flux increase of 
the $\lambda$1032 line itself would result in a decrease of 
the conversion efficiency (Eq.~(3)). 
A comparison of the derived efficiencies to a grid of model 
calculations published by \cite{schmid96} suggests that a kind 
of his $XB3$ model could be relevant, i.e. nearly one third 
of the sky, as seen from the \ion{O}{vi} ionized zone, is 
covered by the neutral scattering region. 

  (iii) 
A significant increase in the Raman conversion efficiency 
during the first quadrature after the spectroscopic conjunction 
(2002 autumn -- 2003 spring, orbital phases 0$\div$0.25, Fig.~8) 
reflects an enlargement of the neutral region seen from the 
\ion{O}{vi} zone. The observed \ion{O}{vi}\,$\lambda$1032 flux 
decreased, while its Raman scattered counterpart at $\lambda$6825 
increased (Table~3). Here the efficiency of 10$\div$12\% are 
similar to that of the $XC3$ model of Schmid's (1996) 
simulations, in which at least one half of the '\ion{O}{vi} sky' 
is occupied by atoms of neutral hydrogen. 
This change in distribution of the neutral material in the 
binary could be connected with an enhancement of the hot 
star wind near the equatorial plane as a result of dilution 
of the optically thick disk-like shell during this period 
(Sect.~4). 
According to hydrodynamical calculations that include effects 
of the orbital motion, a dense S-shaped wind-wind collision 
region can be formed between the stars where the fast wind 
pushes the giant wind \citep[e.g. Fig.~3 of][]{walder}. This 
interaction zone can divide the binary space into two 
approximately equal parts of the neutral and ionized wind. 
Such a distribution of neutral giant's wind in the active binary 
satisfies conditions required by a larger efficiency of the Raman 
scattering process. 

  (vi) 
During the post-outburst period (2003, orbital phase 
$\sim$\,0.5), observed \ion{O}{vi} fluxes increased, while 
the $\lambda1032 \rightarrow \lambda6825$ conversion 
efficiency decreased. This indicates a decrease 
of the size of the scattering region as 'seen' by the 
\ion{O}{vi} photons. Derived efficiencies of $\sim$\,5\,\% 
are similar to those from quiescence \citep{bes98,schmid+99}. 
%
%
%==========================================|
%--- Fig. 8: 1032 ---> 6825 efficuency  ---|
%==========================================|
%
\begin{figure}[p!t]
\centering
\begin{center}
\vspace*{-2mm}
\resizebox{\hsize}{!}{\includegraphics[angle=-90]{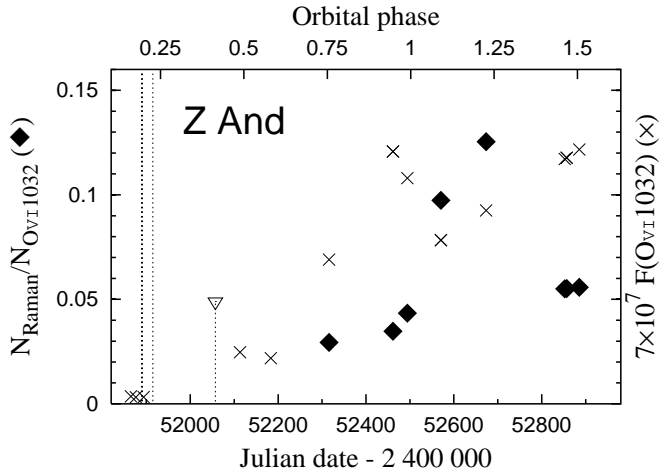}}
\caption[]{Evolution of the Raman conversion efficiency 
          $\lambda$1032$ \rightarrow \lambda$6825 along the outburst 
          of Z\,And (full diamonds). Open triangles connected with 
          dotted line denote upper limits for too low Raman fluxes 
          that were beyond 
          our detection (see the text). Data are from Table~3. 
          }
\end{center}
%\label{fig_1}
\end{figure}

\subsubsection{Comparison of line profiles}
%^^^^^^^^^^^^^^^^^^^^^^^^^^^^^^^^^^^^^^^^^^

Evolution in the Raman scattered \ion{O}{vi}\,$\lambda$1032 
line along the outburst is shown in Fig.~4. A comparison of 
both direct and Raman scattered line is plotted in Fig.~9 
for near-simultaneous optical/far-UV observations. Table~4 
summarizes their basic parameters ($RV_{\rm c}$, FWHM). 
Well-measurable Raman $\lambda$6825 line developed from 
07/08/02 on our spectra, i.e. nearly two years after the 
beginning of the outburst. Profiles consist of a central 
peak at $\sim$\,6832\,\AA\ and a violet and red shoulder at 
$\sim$\,6823\,\AA\ and $\sim$\,6837\,\AA, respectively. 
On 02/02/03 the red adjacent component developed in 
a separate peak of a comparable intensity to the main peak. 
The directly measured \ion{O}{vi}\,$\lambda$1032 line was 
characterized by a strong central emission shifted by 
about +58\kms\ with respect to the systemic velocity. 
Such a large redshift was measured also during 
a quiescent phase by \cite{schmid+99} who explained this 
effect by radiative transfer in strong resonance lines 
throughout an expanding medium. 
During the active phase, this line absorption/scattering 
effect attenuated the violet wing from about $+$50 to 
$-$350\kms\ and produced a P-Cygni type of the profile 
(Fig.~3). This suggests its origin in the densest part of 
the hot star wind. Other variations along the active phase 
(Fig.~9) are discussed below. 

We compare the line profiles in the radial velocity (RV) 
space considering the principle of energy conservation 
of the Raman scattering process. According to \cite{nsv89} 
a broadening $\Delta\lambda_{\rm i}$ of the incident 
line $\lambda_{\rm i}$ is transferred to a broadening 
$\Delta\lambda_{\rm f}$ of the scattered line 
$\lambda_{\rm f}$ as 
\begin{equation}
 \frac{\Delta\lambda_{\rm f}}{\lambda_{\rm f}} = 
 \frac{\Delta\lambda_{\rm i}}{\lambda_{\rm i}} 
 \frac{\lambda_{\rm f}}{\lambda_{\rm i}}.
\end{equation}
In the RV-scale and for the 
\ion{O}{vi}\,($\lambda1032 \rightarrow \lambda6825$) scattering 
process, Eq.~(4) can be written as 
$\Delta RV_{\ion{O}{vi}} = \Delta RV_{\rm Ram}/6.614$. 
Further we assume: 
  (a) The \ion{O}{vi} emission region is connected with the wind 
      from the hot star. 
%so the O$^{+5}$ nebula is concentrated around 
%it and expands in the form of a stellar wind. 
  (b) The scattering region is represented by neutral atoms 
      of hydrogen of the wind from the giant. 
  (c) Between the binary components there is an interaction 
      zone of the two winds. It separates a low density (filled 
      with material from the hot star) and a high density region 
      (filled with material from the giant). The zone can be 
      spiral-shaped due the orbital motion \citep[e.g.][]{walder}. 
With the aid of a general discussion on the \ion{O}{vi} line 
profiles of 6 objects as given by \cite{schmid+99}, we propose 
the following explanation of our observations (Fig.~9): 

 (i) 
The first measured Raman $\lambda$6825 line (10/02/02, 
$\varphi$ = 0.76) was compared with the direct \ion{O}{vi} 
line from 30/09/01 ($\varphi$ = 0.58). Unusually, with respect 
to later observations, both profiles nearly overlapped each 
other. So there were no significant motions of the scattering 
region relative to the direction of the incoming \ion{O}{vi} 
photons. 
This implies that the \ion{O}{vi} photons could not impact 
the neutral wind at the orbital plane moving toward them. 
Otherwise the $\sim-40$\kms\ shift of the Raman line relative 
to the original line had to be observed \citep{schmid+99}. 
According to our interpretation of the hot object structure 
(Sect.~4), a disk-like material blocked the \ion{O}{vi} photons 
at the orbital plane, which thus could 'see' only outer parts 
of the giant's wind giving rise to only a small redshift in the 
$\lambda$6825 profile. %with respect to its initial line. 
Note that the major part of the \ion{O}{vi} flux is created 
in the densest regions of the wind encompassing the hot star 
(cf. Appendix~B). 
Another peculiarity is that the violet wing of the $\lambda$1032 
line was not so strongly absorbed in the wind medium as we 
observed later during the outburst. As a result the total profile 
was shifted only to $\sim+35$\kms\ (Table~4). A shallow 
minimum in the $\lambda$1032 line at $\sim+18$\kms\ could 
be caused by a self-absorption near to the hot object. Note 
that the radial velocity of the hot component orbital motion 
at the time of the observation ($\varphi$ = 0.76) is comparable 
to this shift \citep[parameters from][]{mk96}, which thus 
places its origin at/around the hot star, in agreement with 
our assumption (a) above. 
%
%==================================================|
%----- Table 4: Line parameters: Raman + OVI ------|
%==================================================|
%
\begin{table}
\caption[]{Line center ($RV_{\rm c}$) and line width (FWHM) 
           for direct and Raman scattered 
           \ion{O}{vi}\,$\lambda$1032 line in \kms.
          }
\begin{center}
\begin{tabular}{ccccc}
\hline
\hline
 Date                             & 
 \multicolumn{2}{c}{$RV_{\rm c}$} & 
 \multicolumn{2}{c}{FWHM}         \\
                                  &
 $\lambda$6825                  &
 $\lambda$1032                  &
 $\lambda$6825                  &
 $\lambda$1032                  \\
\hline
%  Date          RVc            FWHM     
%            6825   1032    6825    1032
%--------------------------------------------------
% 16/11/00 &  --   &  xx   &   -   &   yy    \\
% 27/11/00 &  --   &  xx   &   -   &   yy    \\
% 15/12/00 &  --   &  xx   &   -   &   yy    \\
 22/07/01 &  --   &  30   &   -   &  145    \\
 30/09/01 &  --   &  35   &   -   &  150    \\
 10/02/02 &  46   &  --   & 160   &   --    \\
 05/07/02 &  --   &  67   &  --   &   34    \\
 07/08/02 &  36   &  --   & 120   &   --    \\
 22/10/02 &  --   &  64   &  --   &   93    \\
 02/02/03 &  41   &  --   & 139   &   --    \\
 31/07/03 &  33   &  --   & 118   &   --    \\
 04/08/03 &  --   &  59   &  --   &   77    \\
 02/09/03 &  35   &  --   & 120   &   --    \\
%\hline
\multicolumn{5}{c}{Quiescent phase}         \\
 31/01/95$^{\dagger}$ &  --   &  --   & $\approx$113 &  --  \\
 06/03/95$^{\dagger}$ &  --   &  --   &  --  &     70       \\
 28/11/96$^{\ddagger}$ &  33   &  57   & 101   &   78       \\
\hline
\end{tabular}
\end{center}
 ${\dagger}$ -- Birriel et al. (2000), 
${\ddagger}$ -- \cite{schmid+99}
\end{table}
%
%
%==================================================|
%---- Fig. 9: FUSE OVI & Raman line profiles  -----|
%==================================================|
%
\begin{figure}%[p!t]
\centering
\begin{center}
\resizebox{\hsize}{!}{\includegraphics[angle=-90]{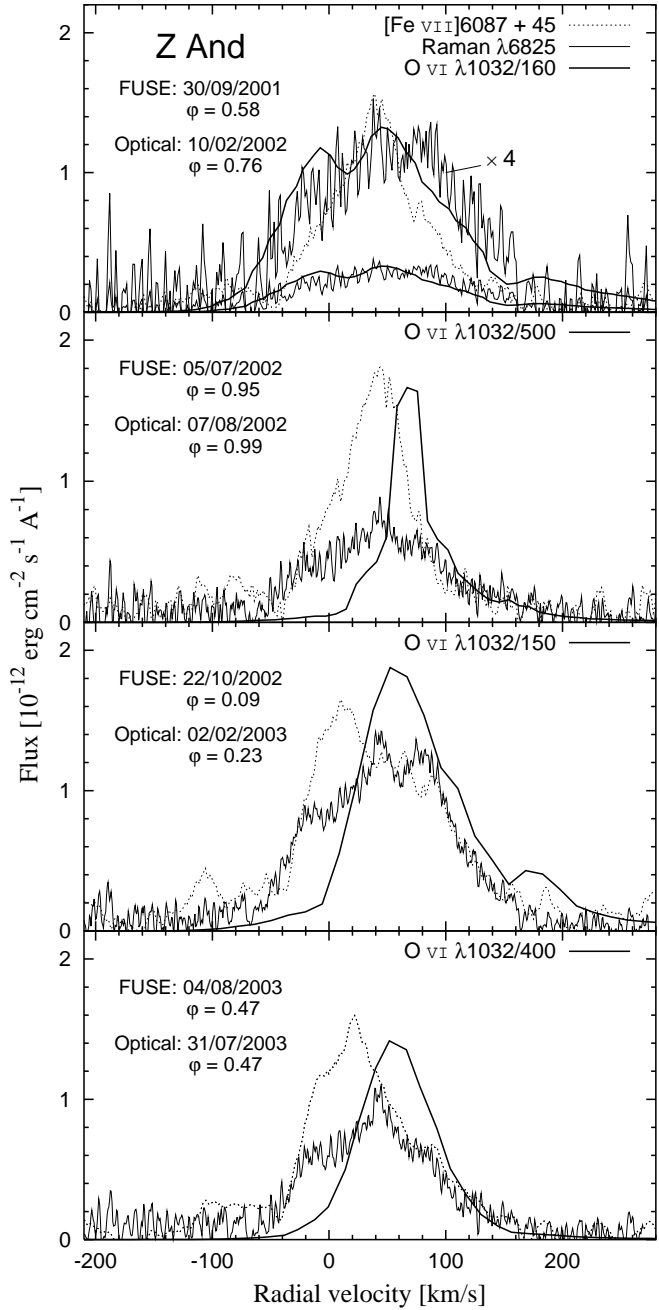}}
%\resizebox{8cm}{!}{\includegraphics[angle=-90]{profiles.eps}}
%
\caption[]{
Comparison of the \ion{O}{vi}\,$\lambda$1032 line profiles 
with their Raman scattered counterparts along the outburst. 
Compared is also the [\ion{Fe}{vii}]\,$\lambda$6087 line 
profile shifted by +45\kms\ (see Sect.~3.4.3). The \ion{O}{vi} 
line flux was scaled to that of the Raman line. 
          }
\end{center}
%\label{fig_1}
\end{figure}

 (ii) 
Next available pair of \ion{O}{vi}\,$\lambda$1032 and 
$\lambda$6825 lines was taken just prior to the inferior 
conjunction of the red giant ($\varphi$ = 0.95 and 0.99). 
The line flux at $\lambda$1032 was the largest from our sample 
and its profile was characterized by a narrow redshifted core 
($RV_{\rm c}\,\sim\,$+67\kms), while its scattered counterpart 
at $\lambda$6825 was rather faint. 
The profile of the initial \ion{O}{vi} line was probably 
modified by an occultation of the \ion{O}{vi} nebula, which 
thus reduced its violet wing for the outer observer. Note 
that from $\varphi \sim 0.94$ a decrease of the optical light 
due to the eclipse effect was observed \citep[Fig.~2 of][]{sk03a}. 
On the other hand, the \ion{O}{vi} zone was not subject 
to occultation for the scattering region spread within 
the \ion{H}{i} zone in the binary. So this could 'see' the 
un-affected \ion{O}{vi}\,$\lambda$1032 line profile and 
thus convert it to a standard type of the $\lambda$6825 
line profile as measured at other orbital phases later on. 
The significant increase of the direct \ion{O}{vi} line flux 
at approximately unchanged or slightly increased conversion 
efficiency constrains also an enhancement of the \ion{H}{i} 
atoms taking part in the scattering process (Eq.~3). 
The former resulted from the temperature increase of the ionizing 
source (Sects.~3.2\,(vii) and 3.3), while the latter could result 
from a dilution of the cooler disk-like pseudophotosphere. 
As a result the wind from the hot star could appear close 
to the orbital plane and thus met a massive neutral wind from 
the giant moving against it. This lead to the appearance of 
the violet adjacent component in the $\lambda$6830 profile 
and an enhancement of the $\eta(\ion{O}{vi})$ efficiency. 

 (iii) 
After the spectroscopic conjunction (the far-UV and optical 
observation made at $\varphi$ = 0.09 and 0.23, respectively; 
Table~3), during the 2002/03 transient increase in the star's 
brightness, we observed the largest efficiency of the 
      $\lambda1032 \rightarrow \lambda6825$ 
transition, well developed three-peaked structure of the Raman 
line and a relatively broad line core of the direct \ion{O}{vi} 
line with superposition of a wind component at $\sim\,+180$\kms\ 
(Fig.~9). The Raman line was broader by the blue adjacent 
component. Its wing was shifted by about --40\kms\ relatively 
to that of the $\lambda$1032 line. 
This emission is produced by the \ion{H}{i} atoms of the giant's 
wind, which are located along the binary axis and move toward
the \ion{O}{vi} region. 
The absorption component in the emission core of the H$\alpha$ 
profile at the position of $\approx-45$\kms\ 
(Sect.~3.2\,(ii)) supports this interpretation. 
In addition, a dense neutral region could be created as 
an enhanced hot star wind pushes the neutral giant's wind 
between the binary components. 
This can increase the effective number of the hydrogen atoms 
and thus also the efficiency $\eta(\ion{O}{vi})$. 
The outer part of the neutral wind and the \ion{H}{i} atoms 
in the collisional zone moving from the incoming \ion{O}{vi} 
photons then extend the red side of the Raman line profile. 
%
%The pronounced violet shoulder in the Raman line profile could
%be a result of an absorption in a dense neutral wind-pushed
%region around the binary axis. 
%
The red wind component in the $\lambda$1032 profile has no 
Raman counterpart. This portion of the \ion{O}{vi} emission 
has an origin at outer part of the hot wind, i.e. further away 
from the scattering region, which reduces the scattering 
efficiency. 
%
%A sketch is drown in Fig.~z. 

 (iv) 
Similar profiles were observed also later at the orbital phase 
$\varphi$ = 0.5 (nearly simultaneous observations, bottom panel 
of Fig.~9). This suggests the difference between the direct 
and the Raman line profile to be of the same nature as we 
discussed above in the point 3. 

The shift between the red wings of the Raman-scattered and the 
direct \ion{O}{vi} line during a quiescent phase was larger 
than we observed during this activity. FWHM of the Raman line 
from the activity exceeded this quantity from quiescence 
by about 10$\div$30\kms, while the width of the incident 
\ion{O}{vi} line did not change so much 
\citep[][Table~4 here]{schmid+99,bes00}. 
This reflects a different kinematics of the scattering region 
during active and quiescent phases. 
Because of a strong wind from the hot star developed during 
the active phase, this could be due to the wind-wind interaction 
zone in the binary. 
Therefore, in the following section, we discuss possible 
connections between profiles of the 
[\ion{Fe}{vii}]\,$\lambda$6087 and Raman\,$\lambda$6825 line. 

\subsubsection{Raman~$\lambda$6825 and 
           {\rm [}\ion{Fe}{vii}{\rm ]}\,$\lambda$6087 profiles}
%^^^^^^^^^^^^^^^^^^^^^^^^^^^^^^^^^^^^^^^^^^^^^^^^^^^^^^^^^^^^^^

A comparison of [\ion{Fe}{vii}]\,$\lambda$6087 and Raman 
$\lambda$6825 line profiles is shown in Fig.~9. 
The former was shifted by +45\kms\ to eliminate the redshift 
in the Raman line (Sect.~3.4.2). 
Similarity of both profiles is obvious, mainly in 2002 
and 2003 when also their line fluxes were comparable. 
Their profiles -- a large FWHM of $\approx$120-140\kms, steep 
wings with a multi-peak structure at the top -- suggest them 
to be due to kinematics of emitting regions. 
Wings of both the lines practically overlap each other at least 
up to the half of the maximum. Striking is the nearly equality 
of the extended red wings with a pattern of the red adjacent 
component (bottom panels of Fig.~9). 

These common signatures of both line profiles suggest 
a similar kinematics for the responsible emitting regions, 
but differing drastically in conditions of ionization -- one 
region has to contain \ion{H}{i} atoms at high density, while 
the other one the Fe$^{+6}$ ions at a low density. This can 
be satisfied if these lines are formed within the wind-wind 
interaction zone. The Raman line can be created at/around 
the interaction surface from the giant's side, where its 
neutral wind is pushed by the hot star wind, while the 
[\ion{Fe}{vii}] emission can arise in a collisionally 
ionized boundary layer between the two winds. 
The observed change in the line width at $\varphi\sim$\,0.25 
(02/02/03, a maximum width) and $\varphi\sim$\,0.5 (31/07/03, 
a narrower width) is consistent with the main wind flow velocity 
aligned approximately perpendicular to the line connecting 
the stars. 
Previous works also support such origin of [\ion{Fe}{vii}] 
lines. \cite{w+84} showed that line profiles of forbidden 
lines from highly ionized atoms could result from interaction 
nebula of wind-wind collision in a binary system. The observed 
profile of the [\ion{Fe}{vii}] line in V1016\,Cyg was very 
similar to our observations from 2003 (see their Fig.~4). 
Also hydrodynamical calculations of the colliding zone of 
the two winds by \cite{nw93} showed a wide variety of 
resulting line profiles, some of which are similar to our 
observations (e.g. the overall broadening with steep sides of 
profiles). 
%
% X - inconsistence with formation in a sph. symm. region X
%
%The radial velocity dispersion and shaping 
%of the [\ion{Fe}{vii}] profiles is inconsistent with their 
%formation in a spherically symmetrical region near to the hot 
%star. Wind velocities of $\sim$\,300, $\sim$\,1200 and 
%$\sim$\,2500\kms\ were indicated by \ion{O}{vi}\,$\lambda$1032, 
%\ion{He}{ii}\,$\lambda$4686 and H$\alpha$ line profiles 
%(Sect.~4), while the material in the interaction zone can move 
%at about 100\kms\ \citep{w+84,gw87}, which satisfies broadening 
%of the [\ion{Fe}{vii}]\,$\lambda$6087 line. 
%However, its quantitative modeling for the present case 
%would be desirable. 
%
% == diagnostic tool:
%
%Properties of these two lines provides an evidence of 
%the wind-wind interaction region between the binary components 
%during the investigated activity of Z\,And. 
% ==
%

Accordingly the observed broadening of the Raman line profile 
together with profiles of highly ionized forbidden lines 
can provide a diagnostic tool in studying motions in 
the wind-wind interaction zone during {\em active phases}. 
Note that a Raman photon may escape easily from the neutral 
region, because this elastic process cannot be repeated 
(in contrast to Rayleigh scattering) and thus reflects 
motions in the densest (a small cross-section) outer layers 
illuminated by \ion{O}{vi} photons. 

\subsection{On the origin of the broad H$\alpha$ wings}
%^^^^^^^^^^^^^^^^^^^^^^^^^^^^^^^^^^^^^^^^^^^^^^^^^^^^^^

\cite{nsv89} pointed out that the broad H$\alpha$ wings in 
symbiotic stars could be formed through Raman scattering of 
Ly$\beta$ photons by atomic hydrogen. \cite{l00} and 
\cite{lh00} elaborated a relevant model. They showed that 
the wing profile formed through the Raman scattering can 
be approximated by the curve proportional to 
$f(\Delta v) \propto \Delta v^{-2}$ in the velocity 
co-ordinates. On the other hand, \cite{sk+02a} derived an 
expression of the same type (see their Eq.~11) to fit 
the broad components of the hydrogen line profiles that form 
in an optically thin and spherically symmetric stellar wind. 
Therefore it is not possible to distinguish the wind emission
from the scattered features in the H$\alpha$ wing directly. 
In addition, the observed Ly$\beta$ profile can be influenced 
by other processes in the circumstellar medium. 

In our case the observed Ly$\beta$ profile is not able to 
produce the H$\alpha$ wings through the Raman scattering 
process. The width of the scattered H$\alpha$ wings is 
proportional to the width of the Ly$\beta$ line as 
$\Delta v_{\rm Ly\beta} = \Delta v_{\rm H\alpha}$/6.4. 
Then the full extension of H$\alpha$ wings $\ga 4\,000$\kms\ 
(Fig.~4) requires $\Delta v_{\rm Ly\beta} > 625$\kms\ or 
$\Delta\lambda_{\rm Ly\beta} > 2.1$\,\AA, which is a factor 
of $\sim$2 larger than the observed 
FWZI(Ly$\beta$)\,$ < 1.0$\,\AA. 
In the case of an optically thick Rayleigh scattering region 
the Ly$\beta$ photons may undergo many Rayleigh scatterings 
until they are converted into Raman photons and escape. 
Thus in such the special case only Raman features can be 
detected, but no original Ly$\beta$ emission. 
The real case is complicated with the presence of a strong 
and variable geocoronal component of Ly$\beta$, which is 
difficult to isolate from the circumstellar one. 
As a result we have to investigate other possibilities 
to identify the main source of radiation contributing to 
the broad H$\alpha$ wings. 
For example, \cite{i+04} found that polarization profile of
H$\alpha$ does not agree with that of the Raman $\lambda$6830
line for Z\,And on 25/10/02 (i.e. during the 2002/03 brightening). 
This finding suggests that the H$\alpha$ line does not include 
the Raman-scattered component that originates from Ly$\beta$. 
Also a relationship between the emission from \ha\ wings 
and the emission measure of the symbiotic nebula supports 
the origin of the broad \ha\ wings in the ionized stellar 
wind \citep[][]{sk06}. 

Therefore we suggest that the extended H$\alpha$ wings, which 
developed during the investigated active phase of Z\,And, are 
due to kinematics of the ionized hydrogen rather than the Raman 
scattered Ly$\beta$ photons. However, this problem deserves 
further investigation. 

\subsection{Optical rebrightening in 2002/03}
%^^^^^^^^^^^^^^^^^^^^^^^^^^^^^^^^^^^^^^^^^^^^
% Observational characteristics: 
%
After 2002 September/October, when the hot object was arising 
up from the eclipse, a rebrightening in the LC 
($\Delta U \sim 0.6$, 
 $\Delta B \sim 0.4$, 
 $\Delta V \sim 0.3$) 
was observed for about 5 months. 
Our estimate of its magnitude increase takes into account 
the depth of the eclipse and the brightness level prior to 
it \citep[][]{sk03a}. 
During this event we measured an additional extension of 
the H$\alpha$ wings on our spectrum from 02/02/03 and 
\cite{t+05} observed such broadening also in the emission 
wings of the \ion{He}{ii}\,$\lambda$4686 line profile with 
FWZI$\sim$2400\kms. 

The observed indices from the maximum (around 15/12/2002) were 
%                    $U_{\rm obs} \dot =  10.05$, 
%                    $B_{\rm obs} \dot =  10.75$, 
%                    $U_{\rm obs} \dot =   9.90$), 
  $(U-B)_{\rm obs}^{\rm max}\,\dot =\,-0.70$ 
and 
  $(B-V)_{\rm obs}^{\rm max}\,\dot =\,0.85$. 
If we deredden these observations with $E(B-V) = 0.30$, correct 
for emission lines (we adopted $\Delta U_{l} = 0.25$, 
                               $\Delta B_{l} = 0.45$ and 
                               $\Delta V_{l} = 0.18$ 
corresponding to a standard emission line spectrum for classical 
symbiotics \citep[][]{sk03b}) 
and subtract contributions from the giant,
  $F_{\rm g}(B) =  1.3\times 10^{-13}$\ecsa
and
  $F_{\rm g}(V) =  2.9\times 10^{-13}$\ecsa\
(see Fig.~7), we get indices of the extra light that causes 
the optical rebrightening as  
  $(U-B)_{\rm extra}^{\rm max} \dot = -1.27$
and
  $(B-V)_{\rm extra}^{\rm max} \dot = 0.59$.
These quantities reflect a large contribution from the nebula 
radiating at high electron temperature \citep[e.g.][]{m+82}. 
We demonstrate this case in Fig.~7. The $UBVR$ flux-points 
from the maximum of the 2002/03 rebrightening suggest the same 
type of the SED as in the mid-2003, but with a factor 
of $\sim$\,2 stronger contribution from the nebula 
($EM({\rm rebrightening}) \approx 2.6\times 10^{60}$\cmt). 

% - Our interpretation 
%
Assuming a particle-bounded nebula (i.e. geometrically open in 
the sense of the STB model), a maximum increase of its radiation 
at a constant flux of ionizing photons can be produced by 
an injection of new particles into it, that 
will convert a relevant amount of the ionizing photons into 
the nebular radiation to yield a radiation-bounded 
(closed) nebula. The particle-bounded nebula 
in Z\,And during 2002-03 is supported by a high quantity of 
the hydrogen ionizing photons of 
$\ga 1.2\times 10^{47}$\,s$^{-1}$ given by the hot object 
luminosity ($L_{\rm h} > 1\,800\,L_{\sun}$) and its temperature 
($T_{\rm h} > 120\,000$\,K, Table~2). A large value of the 
parameter $X~(\approx 14)$ during quiescence corresponding 
to an open nebula was noted by \cite{fc+88}. 
Within this scenario we interpret the small optical rebrightening 
as a consequence of a dilution of the optically thick shell 
around the hot star. This event supplied additional emitters 
into the particle-bounded nebula, which can convert a relevant 
part of the far-UV radiation through $f-b$ and $f-f$ transitions 
to the optical. 
By other words, the material became more transparent and thus 
was subject to ionization by the hot central star even at the 
orbital plane, which led to a transient increase of the nebular 
emission. 
We add that similar brightening was also observed in the LC 
of AX\,Per at its transition from the active to quiescent 
phase \citep[Figs.~3 and 10 of][]{sk+01}. 
A dilution of an optically thick shell in this case was directly 
confirmed by the change of the narrow eclipses into very broad 
minima occurring just after the event (cf. Fig.~3 there). 

% - Sok+05 interpretation
%
\cite{sok+05} interpreted the 2002/03 
rebrightening as a result of a decrease in $T_{\rm h}$ (from 
around 150\,000\,K to approximately 120\,000\,K) due to 
a slight expansion of the white-dwarf photosphere. They 
based this view on that both the $FUSE$ and $VLA$ fluxes 
were unusually low during the optical rebrightening and 
the $U-B$ index varied from about --0.45 to --0.8 from 
the beginning to the end of the event. 
We put following arguments against this interpretation:
 (i) The decrease in the $FUV$ fluxes and $T_{\rm h}$ 
implies a decrease in both the stellar and the nebular component 
of radiation from the hot object. Note that such the small 
temperature decrease at $T_{\rm h} > 100\,000$\,K induces 
a negligible (i.e. not measurable) change in the SED 
\citep[e.g.][]{mb92}. 
So, we should observe a decrease of the optical brightness. 
 (ii) There is an additional attenuation of the UV continuum 
that is well pronounced for eclipsing systems at positions 
around the inferior conjunction of the giant. This extinction 
process was first noticed by \cite{m+91} and \cite{d+99} 
analyzed it for RW\,Hya. \cite{sk05} found this effect 
to be present in all IUE spectra of Z\,And in the range of 
phases $\varphi\,\approx\,0\,\pm\,0.15$. Therefore the 
'unusually' low $FUSE$ fluxes from 22/10/2002 
($\varphi = 0.093$) result from this effect. 
 (iii) The observed fading in the radio fluxes is consistent 
with that in $EM$: 
$EM({\rm maximum})/EM(2002/03) \approx 
F_{\rm 5GHz}({\rm maximum})/F_{\rm 5GHz}(2002/03)$ 
within the corresponding uncertainties 
\citep[Table~2 here and Table~5 of][]{sok+05}. 
 (iv) A redder $U-B$ index at the beginning of the rebrightening 
(--0.45) and a more blue values at/after the maximum (--0.8) 
could qualitatively reflect variation in optical properties 
of the ionized medium during/after the shell dilution. 

\section{Structure of the hot object during the outburst}
%^^^^^^^^^^^^^^^^^^^^^^^^^^^^^^^^^^^^^^^^^^^^^^^^^^^^^^^^
%
%Based on observational characteristics and results of our 
%ana\-ly\-ses we reconstruct the structure of the hot object 
%in Z\,And during its 2000-03 active phase in 
%the following way: 

\subsection{The stage at/around the optical maximum} 

This stage was characterized by: 

(i) The two-temperature-type of the hot object spectrum. 
The cooler one was produced by a relatively warm stellar source 
radiating at 20\,000$\div$30\,000\,K (Fig.~7, Table~2) and the 
hotter one was represented by the highly ionized emission lines 
(e.g. \ion{He}{ii}, \ion{O}{vi}, [\ion{Fe}{vii}]) and a strong 
nebular continuum radiation ($EM \ga 10^{60}$\cmt, Table~2). 
The former is not capable of producing the observed amount of
the nebular emission and thus the latter signals the presence 
of a hot ionizing source ($\ga 10^{5}$\,K) in the system, which 
is not seen directly by the outer observer.  

(ii) Signatures of a mass-outflow at moderate velocities 
    ($\sim$\,100$\div$200\kms), as indicated by the P-Cygni 
    profiles (Sect.~3.2\,(i), Figs.~2 and 6; also 
    Sokoloski et al.~2005), and at very high velocities 
    ($\approx$\,1000$\div$2000\kms) having a pattern in broad 
    emission wings of \ha\ (Fig.~4). 
%
% \ion{He}{ii}\,$\lambda$1084, $\lambda$4686, 
%H$\beta$ and H$\alpha$ (Figs.~2, 4 and 6).

The two-temperature spectrum and the two-velocity type of 
the mass-outflow from the hot object suggest an optically thick 
disk-like structured material surrounding the central hot 
star and expanding at moderate velocities at the orbital plane, 
while at higher latitudes a fast optically thin wind escapes 
the star. 
Due to a high orbital inclination the outer observer can see 
just the optically thick matter of the disk-like shaped warm 
pseudophotosphere, whose outer flared rim occults the central 
ionizing source. Only the surrounding medium from above/below 
the disk can 'see' directly the ionizing source and thus 
convert its radiation to the nebular emission. As a result, 
the temperature of the ionizing source, $T_{\rm h}^{\rm i.s.}$, 
can be obtained only through emission lines 
\citep[][derived $T_{\rm h}^{\rm i.s.}\sim 95\,000$\,K]{sok+05}, 
while the observed UV/optical continuum determines the temperature 
of the stellar source (i.e. the warm pseudophotosphere), 
$T_{\rm h} \sim 27\,000$\,K (Sect.~3.3). Thus this stage 
was characterized by 
\begin{equation}
        T_{\rm h} \ll T_{\rm h}^{\rm i.s.}.
\end{equation}
Also the low fluxes of highly ionized elements, whose profiles 
were shifted blueward (Sect.~3.2\,(iii), Fig.~2 for \ion{He}{ii}, 
Fig.~4 for [\ion{Fe}{vii}]), are consistent with 
the disk-like structure of the hot object. A small size of the 
corresponding ionized zone (see Appendix~B) produced relatively 
weak emission and the optically thick disk-like extended 
source, inclined at $75 - 80^{\circ}$ \citep{sk03a}, blocked 
a fraction of radiation from an outflow at its back side in 
the direction of the observer. As a result the blue-ward shifted 
emission dominated the profile. 
This basic structure of the active object persisted probably 
until about 2002 August, when the eclipse effect in the LC 
was the last indication of the presence of the optically thick
disk-like shell around the hot star \citep{sk03a}. 

Finally, we note that such the structure of the hot object 
is common for all {\em active} S-type symbiotics with a high 
orbital inclination \citep[][Fig.~27]{sk05}. 

\subsection{Evolution after 2002 August}
 
During this period we indicated a more significant dilution 
of the optically thick material around the accretor and 
evolution of a fast wind from the hot star nearby the orbital 
plane. The latter is a direct consequence of the former 
event, because the flux of the hot radiation and particles from 
the accretor was less blocked in directions of the orbital 
plane. These events are supported by the following results: 

 (i) 
A significant increase in the temperature of the stellar 
component of radiation, $T_{\rm h}$, we indicated through 
the model SED (Sect.~3.3.3) and by stronger fluxes of highly 
ionized elements (e.g. Fig.~5). Consequently, we have 
\begin{equation}
         T_{\rm h} \sim T_{\rm h}^{\rm i.s.}, 
\end{equation}
in contrast to the evolution at/around the maximum (Eq.~5). 

 (ii)
The optical rebrightening during 2002.7$\div$2003.3, which led 
to an enhancement of the nebular emission (Sect.~3.6). 

 (iii)
An additional extension of the H$\alpha$ wings on our spectrum 
from 02/02/03. Its temporarily larger terminal velocity and flux 
(Fig.~4) are consistent with the evolution of the hot star wind 
nearby the orbital plane and the enhancement of $EM$ (Sect.~3.6). 
Similar broadening was measured also for 
\ion{He}{ii}\,$\lambda$4686 line by \cite{t+05}. 

 (iv) 
Appearance of a fast wind from the hot star nearby the equator 
plane was also suggested by the [\ion{Fe}{vii}]\,$\lambda$6087 
line profile, which evolved to that arising from the wind-wind 
interaction zone (Sect.~3.4.3). 

 (v)
A sudden increase of the Raman conversion efficiency 
(Sect.~3.4.1, Fig.~8) and the appearance of the pronounced 
violet shifted component in the Raman line profile
(Sect.~3.4.2, Fig.~9). 
Both events need an increase of the scattering acts; the 
latter one also negative velocities between the \ion{H}{i} 
atoms and \ion{O}{vi} photons. 
This can be satisfied if a fast wind from the accretor and 
ionizing photons appear near the orbital plane and thus 
interact with the densest portion of the neutral wind from 
the giant around the binary axis that moves against 
the \ion{O}{vi} photons. 
%
%Both these effects result from interaction of the fast wind 
%and ionizing photons from the accretor with the densest 
%portion of the neutral giant's wind near the orbital plane. 
%
%which reflects a relevant increse of 
%the scattering acts. Assuming a constant mass-loss rate from 
%the giant, this could result from a dilution of the disk's 
%material, because the fast wind and ionizing photons from 
%the accretor then could appear near the orbital plane and thus 
%interact with the densest portion of the neutral giant's wind 
%here. 
%
% (vi)
%The pronounced adjacent component in the Raman line profile 
%that is violet shifted relatively to its original \ion{O}{vi} 
%line (Fig.~9) is a consequence of the point (v) above. Its 
%source is the giant's wind moving towards the \ion{O}{vi} 
%photons at/around the binary axis (Sect.~3.4.2, point (iii)), 
%where the largest relative velocities and density gradient 
%are expected. 
%

Finally, the model SED from 2003 suggests that we still could 
not see just the bare white dwarf, because of a large effective 
hot stellar radius ($R^{\rm eff}_{\rm h} \sim 0.1\,R_{\sun}$, 
Table~2). 

\section{Conclusions}
%^^^^^^^^^^^^^^^^^^^^
We studied the symbiotic prototype Z\,And by using the far-UV, 
optical low- and high-resolution spectroscopy and $UBVR$ 
broad-band photometry carried out along its major 2000-03 
outburst. We summarize the main results as follows: 

(i)
An expansion of the hot object and an enhanced mass loss from 
the system was indicated by the change in the SED during the 
transition from the second plateau stage to the maximum 
(Sect.~3.3.2). 
A complex mass-outflow from the system was recognized in 
the structure of the [\ion{O}{iii}]\,$\lambda$5007 profiles 
(ejection of blobs, creation of adjacent emission components; 
Sect.~3.2, Fig.~2) and of the absorption components in 
\ion{C}{iii} and \ion{P}{v} lines in the $FUSE$ spectra. 

(ii)
Evolution of the two-temperature type of the spectrum at the 
initial stages of the outburst having signatures of 
a mass-outflow at moderate and high velocities suggested 
that the active object consists of an optically thick slowly 
expanding disk-like structured material encompassing the white 
dwarf at the orbital plane and of a fast optically thin wind 
escaping the star at higher latitudes (Sec.~4.1). 
The optically thick disk-like shell persisted probably until 
2002 August. Afterwards we indicated its dilution and 
evolution of a fast hot star wind concentrated more at the 
orbital plane (Sect.~4.2). 

(iii)
A striking similarity of [\ion{Fe}{vii}]\,$\lambda$6087 
and Raman $\lambda$6825 profiles at/after the dilution of 
the disk suggested their origin within the interaction zone, 
where the winds from both the stars collide (Sect.~3.4.3). 
This unusual symbiosis of these lines and properties of the 
Raman scattering process can be helping to study streaming 
along the zone of interaction between the two winds. 

\begin{acknowledgements}

The authors thank the anonymous referee for several helpful comments.
Hans Martin Schmid is thanked for useful discussion about Raman 
scattering process. 
M.W. thanks M. \v{S}lechta for initial reduction of the spectra 
taken at the Ond\v{r}ejov Observatory. 
This research was in part supported by a NATO Science Programme  
within its sub-programme {\scriptsize EXPERT VISIT} (A.S.), by 
a grant of the Slovak Academy of Sciences No.~2/4014/04 and 
the Grant Agency of the Czech Republic, grant No.~205/04/2063. 
A.S. acknowledges the hospitality of the Capodimonte Astrophysical
Observatory in Naples. 
\end{acknowledgements}
%
%%%%%%%%%%%%%%%%%%%%%%%%%%%%%%%%%%%%%%%%%%%%%%%%%%%%%%%%%%%%%%%%%%%%%
%
\appendix
\section{\ion{He}{i} flux ratios}

To predict flux ratios of \ion{He}{i} emission lines, 
$F(7065)/F(5876)$ and $F(6678)/F(5876)$, 
\cite{proga+94} solved the statistical 
equilibrium equation for a given level in their (19 level) 
\ion{He}{i} atom including the radiative and three-body 
recombination and effects of collisional ionization and 
collisional transfer between other levels of the model atom. 
The model requires as input parameters the electron temperature, 
$T_{\rm e}$, the electron density, $N_{\rm e}$, the rate of 
photons capable of ionizing \ion{He}{i} atom and the optical 
depth in the \ion{He}{i}\,$\lambda$3889 line, $\tau_{3889}$, 
as a measure of the population of the metastable $2^3\,S$ level. 
Modeled and observed ratios are shown in Fig.~A.1. 
The location of the flux ratios from the optical maximum in 
2000 December suggests that the \ion{He}{+} emitting region 
is very dense, $N_{\rm e} \ga 10^{12}$\cmt, and radiates at 
$T_{\rm e} \ga 20\,000$\,K. Comparing the observed luminosity 
in the $\lambda$5876 line, $4\pi d^2 F(5876)$, with that 
predicted by its emissivity, $N_{\rm e}^2\,\varepsilon_{5876}\,V$, 
we can determine the effective radius of a spherical nebula as 
\begin{equation}
 R_{\rm eff}(\ion{He}{+}) = \Big(\frac{3\,F(5876)\,d^2}
                    {N_{\rm e}^2\,\varepsilon_{5876}}\Big)^{1/3},
\end{equation}
where the emission coefficient 
$\varepsilon_{5876} \dot = 1\times 10^{-26}$\eps\ 
for 
  $N_{\rm e}=10^{12}$\cmt\ 
and 
  $\tau_{3889}=100$ \citep{proga+94}. 
Then $F(5876) = 73\times 10^{-13}$\ecs\ (Table~A.1), 
and $d$ = 1.5\,kpc yield 
$R_{\rm eff}(\ion{He}{+})\,\dot =\,5\,R_{\sun}$. 

During the recent major outburst \ion{He}{i} fluxes evolved 
toward lower $F(7065)/F(5876)$, whereas during the smaller 
1986-eruption the fluxes evolved toward lower 
$F(6678)/F(5876)$ at nearly constant $F(7065)/F(5876)$ 
(Fig.~A.1). 
In general, this reflects different properties of the nebula, 
which develops during outbursts around the hot active object. 
Qualitatively, during strong outbursts the velocity field of 
the hot star wind can make large differences in the density 
throughout the nebula. As a result, contributions from the 
outer, low-density part of the nebula will enhance more the 
$\lambda$5876 line relatively to other lines and thus make 
lower the considered flux ratios. According to \cite{smits} 
the relative fluxes $F(5876)/F(6678)/F(7065)$ = 2.519/0.682/0.637 
for $N_{\rm e} = 10^{6}$\cmt\ and $T_{\rm e}$ = 20\,000\,K 
(open circle in Fig.~A.1). 
Therefore the real size of the \ion{He}{+} nebula will probably 
exceed the effective radius derived for a constant density 
distribution. 
In addition, its shape is not spherical due to the presence 
of the optically thick disk-like matter at the orbital plane, 
which prevents ionization in these directions (Sect.~4). 
%
%===================================|
%---- Fig. A.1: HeI flux ratios ----|
%===================================|
% [p!th] [!ht]
%
\begin{figure}
\centering
\begin{center}
\resizebox{8cm}{!}{\includegraphics[angle=-90]{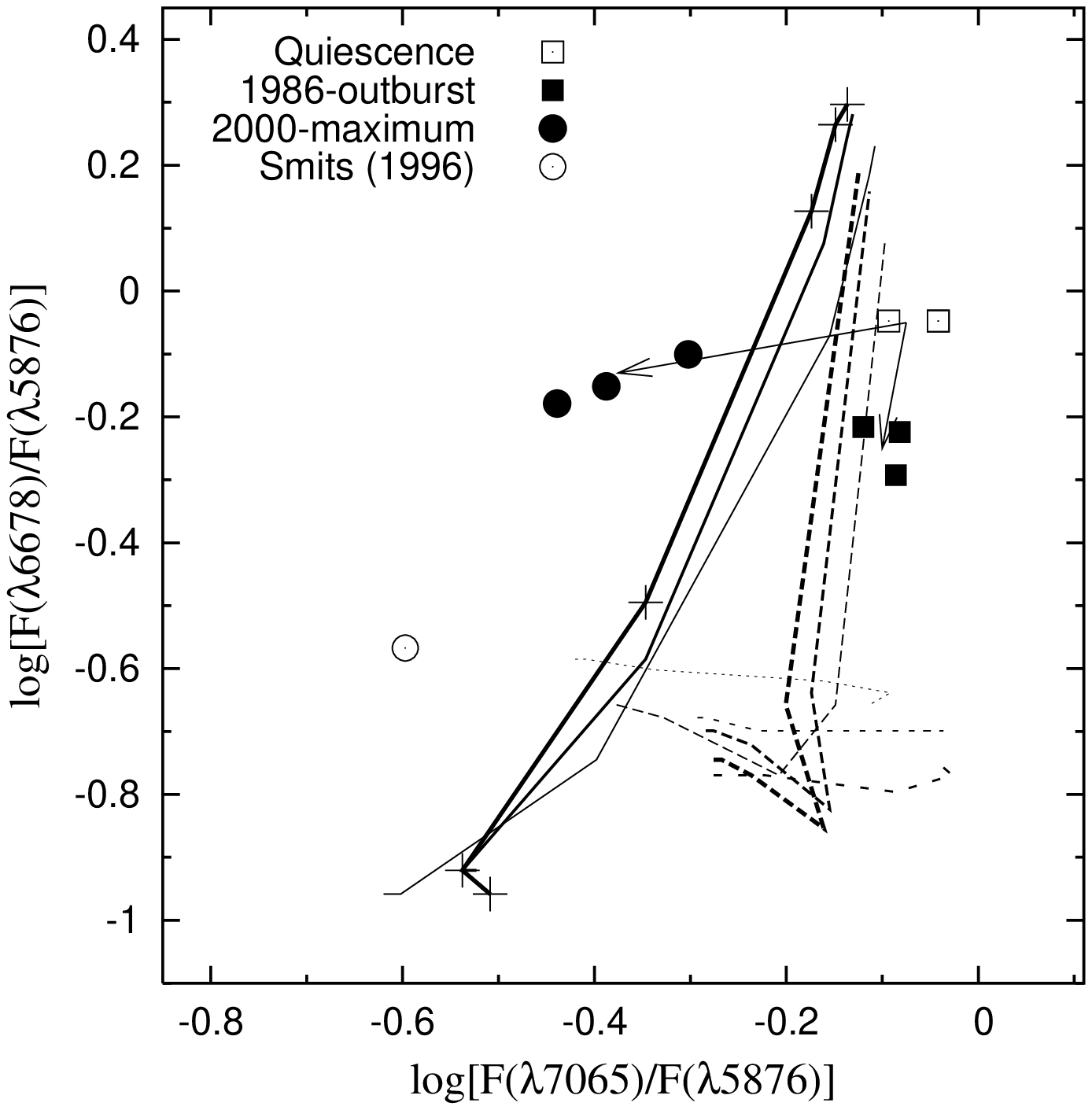}}
%\resizebox{\hsize}{!}{\includegraphics[angle=-90]{hei-r.eps}}
%
\caption[]{
Dependencies of \ion{He}{i} flux ratios, $F(7065)/F(5876)$ 
and $F(6678)/F(5876)$, for $N_{\rm e} = 10^{6}$\cmt\ (dotted 
lines), $N_{\rm e} = 10^{10}$\cmt\ (dashed lines) and 
$N_{\rm e} = 10^{12}$\cmt\ (solid lines), for 
$T_{\rm e}$ = 10\,000\,K, 15\,000\,K and 20\,000\,K (light, 
medium and heavy lines, respectively). Crosses indicate 
$\tau_{3889}$ = 0.01, 0.1, 1, 10, 100, 1000 from the bottom to 
the top on the solid heavy line \citep[Table~5 of][]{proga+94}. 
Observed fluxes are from Table~A.1. 
Open circle corresponds to a low-density nebula 
($N_{\rm e} = 10^{6}$\cmt, $T_{\rm e}$ = 20\,000\,K). 
          }
\end{center}
%\label{fig_1}
\end{figure}
%
%
%==================================|
%----- Table A.1: HeI fluxes ------|
%==================================|
%
\begin{table}
\caption[]{Observed fluxes of \ion{He}{i} emission lines}
\begin{center}
\begin{tabular}{clcccc}
\hline
\hline
 Date & JD~2\,4... & $F(5876)$ & $F(6678)$ & $F(7065)$ & A/Q$^{\dagger}$\\
      &            & \multicolumn{3}{c}{[\ecs]}        &                \\
\hline
 10/12/00 & 51889.4&76.7& 60.4& 36.9 & A \\
 11/12/00 & 51890.4&63.7& 56.4& 37.3 & A \\
 12/12/00 & 51891.3&77.4& 57.2& 33.1 & A \\
 30/09/82 & 45213$^{\ddagger}$ &19 & 19 & 18  & Q \\
 01/06/85 & 46218$^{\ddagger}$ &28 & 19 & 25  & A \\
 02/12/85 & 46402$^{\ddagger}$ &58 & 33 & 56  & A \\
 20/10/86 & 46724$^{\ddagger}$ &39 & 26 & 38  & A \\
 01/12/88 & 47497$^{\ddagger}$ &15 & 15 & 16  & Q \\
\hline
\end{tabular}
\end{center}
$^{\dagger}$ A/Q: Active/Quiescent phase \\
$^{\ddagger}$ from \cite{proga+94}
\end{table}

\section{Radius of the \ion{O}{vi} zone}

Radius of the \ion{O}{vi} zone, $R(\ion{O}{vi})$, can be obtained 
from the equation of the equilibrium between the flux of photons, 
$L_{\rm ph}(\ion{O}{v})$, capable of ionizing \ion{O}{v} ions and 
the number of recombinations by \ion{O}{vi} ions. For a spherically 
symmetric nebular medium around the central ionizing source we can 
write 
\begin{equation}
  L_{\rm ph}(\ion{O}{v}) = \frac{4\pi}{3} R(\ion{O}{vi})^{3}
    A(\ion{O}{vi})N_{+}N_{\rm e}\alpha_{\rm B}(\ion{O}{vi}), 
\end{equation}
where $A(\ion{O}{vi})$ is the abundance of \ion{O}{vi} ions, 
$\alpha_{\rm B}(\ion{O}{vi})$ is the recombination coefficient 
of a free electron with the \ion{O}{vi} ion for the case $B$ 
and $N_{+}$ and 
$N_{\rm e}$ are concentrations of protons and electrons. 
In determining $L_{\rm ph}(\ion{O}{v})$ values we integrated 
the Plank curve from the ionization limit of 113.896\,eV 
($\nu_0 = 27.54 \times 10^{15}$\,s$^{-1}$) to $\infty$. Figure~B.1 
shows the $L_{\rm ph}(\ion{O}{v})$ quantity as a function of the 
temperature for the luminosity of the ionizing source of 
1\,000\,$L_{\sun}$. Further we assumed that all oxygen atoms are 
ionized to \ion{O}{vi} within the zone, i.e. 
$A(\ion{O}{vi}) = A(\ion{O}{i})$, 
for which we adopted the solar photospheric abundance of 
$4.6\times 10^{-4}$ \citep{aspl} and 
$\alpha_{\rm B}(\ion{O}{vi}) = 
        9.5\times 10^{-12}$\,cm$^{+3}$\,s$^{-1}$ 
for $T_{\rm e} = 30\,000$\,K \citep{gur}. 
According to \cite{sok+05}, the temperature of the ionizing source 
$T_{\rm h}^{i.s.}$ = 95\,000\,K and 160\,000\,K during the optical 
maximum and the 2003 post-outburst stage, which implies 
$L_{\rm ph}(\ion{O}{v}) = 6.4\times 10^{43}$ and 
2.6$\times 10^{45}$ photons\,s$^{-1}$ for luminosities of 
7\,000 and 4\,400\,$L_{\sun}$, respectively (Fig.~B.1, Table~2). 
Finally, for the particle concentration of the order of 
$10^{12}$\cmt, which can be expected in a close vicinity of 
the ionizing source due to the mass loss via the stellar wind 
at rates of a few $\times 10^{-7}\,M_{\sun}\,{\rm yr^{-1}}$ 
\citep{sk+06}, one can derive 
%
%
%=================================|
%---- Fig. B.1: L_ph for OVI  ----|
%=================================|
% [p!th] [!ht]
%
\begin{figure}
\centering
\begin{center}
\resizebox{\hsize}{!}{\includegraphics[angle=-90]{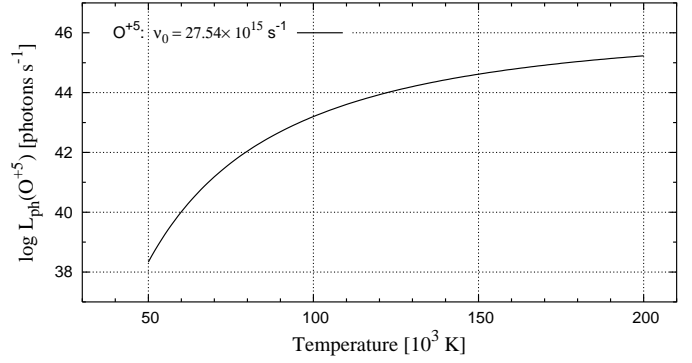}}
\caption[]{
Number of photons capable of ionizing \ion{O}{v} ions 
as a function of temperature scaled to the luminosity of 
1\,000\,$L_{\sun}$. 
          }
\end{center}
%\label{fig_1}
\end{figure}
%
%
%
%==============================================|
%-- Fig. B.2: R(OVI) as a function L_ph(OVI) --|
%==============================================|
% [p!th] [!ht]
%
\begin{figure}
\centering
\begin{center}
\resizebox{\hsize}{!}{\includegraphics[angle=-90]{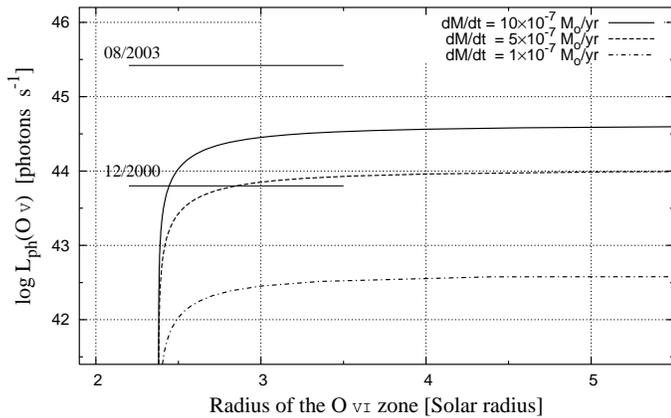}}
\caption[]{
Radius of the \ion{O}{vi} zone as a function of 
$L_{\rm ph}(\ion{O}{v})$ photons for particle distribution 
in a form of the stellar wind from the ionizing source 
(Eq.~B.4). Horizontal bars indicate quantities from the 
optical maximum (2000 December) and the 2003 post-outburst 
stage. 
          }
\end{center}
%\label{fig_1}
\end{figure}
\begin{equation}
   R(\ion{O}{vi}) ~\approx ~{\rm a ~few} ~\times ~R_{\sun} 
\end{equation}
during the optical maximum. In a more rigorous approach we 
integrate densities of the hot 
star wind from a radius $r_1$ above its origin, $R_{\rm w}$, 
to the end of the \ion{O}{vi} zone 
(i.e. the $R(\ion{O}{vi})$ radius). 
Then Eq.~(B.1) takes the form 
\begin{equation}
  L_{\rm ph}(\ion{O}{v}) = 4\pi\,\alpha_{\rm B}(\ion{O}{vi})\,
                           A(\ion{O}{vi})\!
             \int_{r1}^{R(\ion{O}{vi})}\!\!\! N(r)^2\,r^2 {\rm d}r, 
\end{equation}
where $N(r) = \dot M/4\pi r^{2}\mu m_{\rm H} v(r)$ is determined 
via the continuous mass loss rate $\dot M$ from the hot star with 
the velocity field 
$v(r) = v_{\infty}(1 - R_{\rm w}/r)^{\beta}$ \citep{cak}. 
This allows to express the last equation as 
\begin{equation}
  L_{\rm ph}(\ion{O}{v}) = 
          \frac{\alpha_{\rm B}(\ion{O}{vi})\,A(\ion{O}{vi})}
               {4\pi(\mu m_{\rm H})^2}\,
          \Big(\frac{\dot M}{v_{\infty}}\Big)^{2}
          \frac{1}{R_{\rm w}}\times I,
\end{equation}
where the integral 
\begin{equation}
  I = \frac{1}{(2\beta-1)}\,
      \Big[\Big(1-\frac{R_{\rm w}}{r_1}\Big)^{1-2\beta} - 
           \Big(1-\frac{R_{\rm w}}{R(\ion{O}{vi})}\Big)^{1-2\beta}\Big]. 
\end{equation}
Solution of Eq.~(B.4) is shown in Fig.~B.2 for three different 
$\dot M$. Parameters $\beta$ = 1.7, $v_{\infty}$ = 2\,000\kms, 
$R_{\rm w} = 1.7\,R_{\sun}$ and $r_1 \sim 1.4\,R_{\rm w}$ were 
estimated from the modeling the broad H$\alpha$ wings during 
the maximum \citep{sk06}. Also in this case the $R(\ion{O}{vi})$ 
radius can be small, $<3\,R_{\sun}$, for mass-loss rates between 
1\,$\times 10^{-6}$ and 5$\times 10^{-7}$\myr\ and the observed 
flux of ionizing photons 
$L_{\rm ph}(\ion{O}{v}) = 6.4\times 10^{43}$\,s$^{-1}$. 
However, for 
$L_{\rm ph}(\ion{O}{v}) \ga 10^{44}-10^{45}$\,s$^{-1}$ 
the \ion{O}{vi} zone can expand to infinity. This is in 
a qualitative agreement with the observed significant and fast 
increase in the \ion{O}{vi}\,$\lambda$1032, 1038 fluxes 
from the optical maximum (2000 December) to the next $FUSE$ 
observation in 2001 July (Table~3), which thus can be understood 
as a result of the increase in $T_{\rm h}^{i.s.}$ to 
120\,000--150\,000\,K at approximately constant 
luminosity \citep{sok+05}. 
%
%  ----- R E F E R E N C E S -------
%

%
%\newpage

\end{document}